\documentclass[longauth]{aa}
\pdfoutput=1
\usepackage{txfonts}
\usepackage{graphicx}
\usepackage{natbib}
\usepackage{xspace}
\bibpunct{(}{)}{;}{a}{}{,}

\newcommand{\hmpc}{$\,{\rm h}^{-1}$ Mpc\xspace}
\newcommand{\wprp}{$w_p(r_p)$\xspace}

\newcommand{\cmocks}{\textsc{Cmocks}\xspace}
\newcommand{\omocks}{\textsc{Omocks}\xspace}

\begin{document}

\title{Comparison of the VIMOS-VLT Deep Survey with the Munich semi-analytical
  model \thanks{Based on data obtained with the European Southern Observatory
    Very Large Telescope, Paranal, Chile, program 070.A-9007(A), and on data
    obtained at the Canada-France-Hawaii Telescope, operated by the CNRS of
    France, CNRC in Canada and the University of Hawaii. This work is based in
    part on data products produced at TERAPIX and the Canadian Astronomy Data
    Centre as part of the Canada-France-Hawaii Telescope Legacy Survey, a
    collaborative project of NRC and CNRS.}}

\subtitle{I. Magnitude counts, redshift distribution, colour bimodality, and
  galaxy clustering}

\titlerunning{Comparison of VVDS with Munich semi-analytical model
  high-redshift galaxy properties}

\author{
  S. de la Torre \inst{1,2}
  \and B. Meneux \inst{3,4}
  \and G. De Lucia \inst{5}
  \and J. Blaizot \inst{6}
  \and O. Le F\`evre \inst{7}
  \and B. Garilli \inst{2}
  \and O. Cucciati \inst{7}
  \and Y. Mellier \inst{8}
  \and A. Pollo \inst{9,10}
  \and D. Bottini \inst{2}
  \and V. Le Brun \inst{7}
  \and D. Maccagni \inst{2}
  \and M. Scodeggio \inst{2}
  \and L. Tresse \inst{7}
  \and G. Vettolani \inst{11}
  \and A. Zanichelli \inst{11}
  \and C. Adami \inst{7}
  \and S. Arnouts \inst{12,7}
  \and S. Bardelli \inst{13}
  \and M. Bolzonella \inst{13}
  \and A. Cappi \inst{13}
  \and S. Charlot \inst{8}
  \and P. Ciliegi \inst{13}  
  \and T. Contini \inst{14}
  \and S. Foucaud \inst{15}
  \and P. Franzetti \inst{2}
  \and I. Gavignaud \inst{16}
  \and L. Guzzo \inst{1}
  \and O. Ilbert \inst{7}
  \and A. Iovino \inst{1}
  \and H.J. McCracken \inst{8}
  \and C. Marinoni \inst{17}
  \and A. Mazure \inst{7}
  \and R. Merighi \inst{13} 
  \and S. Paltani \inst{18}
  \and R. Pell\'o \inst{14}
  \and L. Pozzetti \inst{13} 
  \and D. Vergani \inst{13}
  \and G. Zamorani \inst{13} 
  \and E. Zucca \inst{13}
}

\authorrunning{S. de la Torre et al.}

\offprints{\mbox{S.~de~la~Torre}, \email{sylvain.delatorre@brera.inaf.it}}

\institute{
  INAF -- Osservatorio Astronomico di Brera, Via Bianchi 46, 23807 Merate, Italy
  \and INAF -- Istituto di Astrofisica Spaziale e Fisica Cosmica di Milano, Via Bassini 15, 20133 Milano, Italy
  \and Max Planck Institut f\"ur Extraterrestrische Physik, Giessenbachstrasse 1, 85748 Garching bei M\"unchen, Germany
  \and Universit\"atssternwarte M\"unchen, Scheinerstrasse 1, 81679 M\"unchen, Germany
  \and INAF -- Osservatorio Astronomico di Trieste, Via Tiepolo 11, 34131 Trieste, Italy
  \and Centre de Recherche Astrophysique de Lyon, UMR 5574, Universit\'e Claude Bernard Lyon-\'Ecole Normale Sup\'erieure de Lyon-CNRS, 69230 Saint-Genis Laval, France
  \and Laboratoire d'Astrophysique de Marseille, UMR 6110, CNRS-Universit\'e de Provence, 38 rue Frederic Joliot-Curie, 13388 Marseille, France
  \and Institut d'Astrophysique de Paris, UMR 7095, Universit\'e Pierre et Marie Curie, 98 bis bd Arago, 75014 Paris, France
  \and The Andrzej Soltan Institute for Nuclear Studies, ul. Hoza 69, 00-681 Warszawa, Poland
  \and Astronomical Observatory of the Jagiellonian University, ul Orla 171, PL-30-244, Krak\'ow, Poland
  \and IRA-INAF, Via Gobetti 101, 40129 Bologna, Italy
  \and Canada France Hawaii Telescope corporation, Mamalahoa Hwy, Kamuela, HI-96743, USA
  \and INAF -- Osservatorio Astronomico di Bologna, Via Ranzani 1, 40127 Bologna, Italy
  \and Laboratoire d'Astrophysique de Toulouse-Tarbes, UMR 5572, CNRS-Universit\'e de Toulouse, 14 av E.Belin, 31400 Toulouse, France
  \and Department of Earth Sciences, National Taiwan Normal University, 88 Tingzhou Road, Sec. 4, Taipei 11677, Taiwan, China
  \and Departamento de Ciencias Fisicas, Facultad de Ingenieria, Universidad Andres Bello, Santiago, Chile
  \and Centre de Physique Th\'eorique, UMR 6207, CNRS-Universit\'e de Provence, 13288 Marseille, France
  \and ISDC, Geneva Observatory, University of Geneva, ch. d'\`Ecogia 16, 1290 Versoix, Switzerland
}

\date{Received 6 August 2010 / Accepted 2 October 2010}

\abstract
% context heading (optional)
{}
% aims heading (mandatory)
{This paper presents a detailed comparison between high-redshift observations
  from the VIMOS-VLT Deep Survey (VVDS) and predictions from the Munich
  semi-analytical model of galaxy formation. In particular, we focus this
  analysis on the magnitude, redshift, and colour distributions of galaxies,
  as well as their clustering properties.}
% methods heading (mandatory)
{We constructed 100 quasi-independent mock catalogues, using the output of the
  semi-analytical model presented in De Lucia \& Blaizot (2007). We then
  applied the same observational selection function of the VVDS-Deep survey,
  so as to carry out a fair comparison between models and observations.}
% results heading (mandatory)
{We find that the semi-analytical model reproduces well the magnitude counts
  in the optical bands. It tends, however, to overpredict the abundance of
  faint red galaxies, in particular in the $i'$~and $z'$~bands. Model galaxies
  exhibit a colour bimodality that is only in qualitative agreement with the
  data. In particular, we find that the model tends to overpredict the number
  of red galaxies at low redshift and of blue galaxies at all redshifts probed
  by VVDS-Deep observations, although a large fraction of the bluest observed
  galaxies is absent from the model. In addition, the model overpredicts by
  about 14 per cent the number of galaxies observed at $0.2<z<1$ with
  $I_{AB}<24$. When comparing the galaxy clustering properties, we find that
  model galaxies are more strongly clustered than observed ones at all
  redshift from $z=0.2$ to $z=2$, with the difference being less significant
  above $z\simeq1$. When splitting the samples into red and blue galaxies, we
  find that the observed clustering of blue galaxies is well reproduced by the
  model, while red model galaxies are much more clustered than observed ones,
  being principally responsible for the strong global clustering found in the
  model.}
% conclusions heading (optional)
{Our results show that the discrepancies between Munich semi-analytical model
  predictions and VVDS-Deep observations, particularly in the galaxy colour
  distribution and clustering, can be explained to a large extend by an
  overabundance of satellite galaxies, mostly located in the red peak of the
  colour bimodality predicted by the model.}


\maketitle

\section{Introduction}

Most accurate cosmological probes tend to favour a flat $\Lambda CDM$
cosmological model and strengthen the hierarchical growth of structure
scenario \citep[e.g.][]{riess98,spergel03,tegmark04,komatsu09}. In this
picture, it is believed that the onset of galaxy formation arises inside dark
matter haloes and that the cosmic history of galaxies follows the hierarchical
evolution of haloes.

Despite the undeniable successes of the $\Lambda CDM$ model to explain a broad
variety of astrophysical observations, the description of stellar mass
assembly and star formation activity in this framework and its confrontation
to observations remain challenging \citep[e.g.][]{delucia09}. The most recent
deep galaxy spectroscopic surveys \citep[e.g.][]{davis03,dickinson03,
  lefevre05a,lilly07}, that enabled the census of galaxy properties up to
$z\simeq2$, have confirmed that the so-called \emph{downsizing scenario}
\citep[][]{gavazzi96,cowie96}, is a primary feature of galaxy formation. About
half of the stellar mass in massive galaxies observed at the present time was
already in place at $z\simeq1$
\citep[e.g.][]{bundy05,cimatti06,arnouts07,ilbert10}, while the number density
of less massive galaxies continues to rise with cosmic time even for redshifts
below unity. Most massive galaxies cease to efficiently produce new stars
earlier than lower mass galaxies, which keep producing stars until very recent
times \citep[e.g.][]{tresse07}.  This scenario is also supported by most
recent observations of the VIMOS-VLT Deep Survey \citep[VVDS,][]{lefevre05a},
one of the largest spectroscopic surveys of distant galaxies.  In particular,
the VVDS-Deep sample provides one with a description of the high-redshift
Universe from $z\simeq0.2$ to $z\simeq2$, where the evolution of the galaxy
luminosity function \citep{ilbert05,zucca06}, galaxy stellar mass function
\citep{pozzetti07}, star formation rate \citep{tresse07,bardelli09}, colour
bimodality \citep{franzetti07}, and galaxy clustering \citep{lefevre05b}
cannot be only explained by a simple scenario of hierarchical growth of
baryons.

The complex physical processes involving baryons on galactic scales likely
explain why the physics of stellar mass assembly and star formation is still
puzzling in the $\Lambda CDM$ model.  Fortunately, large N-body simulations
coupled with semi-analytical treatments of baryonic processes or fully
hydrodynamical simulations have become sufficiently sophisticated to draw
detailed and reliable predictions for galaxy properties
\citep[e.g.][]{weinberg04,springel05,kim09sim}.  Current semi-analytical
models can be tuned to optimally reproduce some of the basic galaxy properties
observed in the local Universe such as the galaxy luminosity function or the
Tully-Fisher relation. Galaxy mock samples constructed from these models are
widely used for a large variety of purposes. However, it is not yet
demonstrated that numerical simulations together with semi-analytical model
can successfully describe galaxy properties at high redshift and the physics
driving the evolution of the various galaxy populations.  There have been few
detailed comparisons made between their predicted galaxy properties and a
broad range of observational measurements at high redshift. Most comparisons
were focused on single observations that has provided important clues, but
still a limited perspective on the global galaxy evolution scenario.

\citet{bower06} compare the $K$-band luminosity function, galaxy stellar mass
function, and cosmic star formation rate from the Durham model \citep{bower06}
with high-redshift observations.  They find that their model match the
observed mass and luminosity functions reasonably well up to
$z\simeq1$. \citet{kitzbichler07} compare the magnitude counts in $B,R,I,K$
bands, redshift distributions for $K$-band selected samples, $B$- and $K$-band
luminosity functions, and galaxy stellar mass function from the Munich model
\citep{croton06,delucia07} with deep surveys measurements. They find that the
agreement of the Munich model with high-redshift observations is slightly
worse than that found for the Durham model. In particular, they find
that the Munich model tends to systematically overestimate the abundance of
relatively massive galaxies at high redshift. They note, however, a
non-negligible dispersion between various observations.

The predicted galaxy clustering properties by semi-analytical models have been
mainly compared with local observations
\citep[e.g.][]{springel05,li06,li07,kim09}. Similar conclusions are obtained
when confronting models to 2dFGRS \citep{colless01} and SDSS \citep{york00}
measurements: semi-analytical models are able to reproduce the overall
observed clustering properties, but fail to predict the measurements in
details, e.g. the different clustering of red and blue galaxies
\citep{springel05}. For the high-redshift Universe, only few data-model
comparisons have been carried out
\citep{mccracken07,coil08,meneux08,meneux09}. These studies highlight a number
of discrepancies between model predictions and observations. In particular,
\citet{coil08} show that red model galaxies are more strongly clustered than
observed at $z\simeq1$, particularly on small scales. Blue galaxies in the
model show instead a lower clustering strength than observed, suggesting a
significant deficit of blue satellite galaxies in the semi-analytical model of
\citet{croton06}. They interpret these discrepancies as due to an incorrect
modelling of the colours of satellite galaxies at high redshift.

In this paper, we compare the basic galaxy properties observed in the
VVDS-Deep spectroscopic sample at $0.2<z<2$ to the outputs of the Munich
model. This is the first of a series of papers that aim at a detailed
comparison between model predictions and a complete set of galaxy properties
and their evolution with cosmic time, derived from the VVDS-Deep
observations. In this first paper, we focus on the magnitude counts, redshift
distribution, colour bimodality, $B$-band and $I$-band luminosity
distributions, and the global clustering measurements. The detailed dependence
of galaxy clustering on luminosity and intrinsic colour is the subject of a
forthcoming paper \citep[][hereafter paper II]{paper2}.

In Sect. 2 we summarise the basic characteristics of the VVDS-Deep sample and
the Munich semi-analytical model. In Sect. 3 we compare the observed magnitude
counts, redshift distribution, colour bimodality, $B$-band, and $I$-band
luminosity distributions with model predictions. In Sect. 4 we present the
galaxy clustering comparison and we discuss our results in Sect. 5. Throughout
this paper we assume a flat $\Lambda CDM$ cosmology with $\Omega_M=0.25$,
$\Omega_\Lambda=0.75$ and $H_0=100~\rm{h~km \cdot s^{-1} \cdot Mpc^{-1}}$. All
magnitudes are quoted in the AB system and for simplicity we denote the
absolute magnitude $M_x-5\log{(\rm{h})}$ as $M_x$.

\section{VVDS-Deep sample and Munich galaxy formation model}

\subsection{The VVDS-Deep spectroscopic sample}

The VIMOS-VLT Deep Survey \citep[VVDS,][]{lefevre05a} is a deep spectroscopic
survey of galaxies performed with the VIsible Multi-Object Spectrograph
\citep[VIMOS,][]{lefevre03} at the European Southern Observatory's Very Large
Telescope (ESO--VLT). The deep part of the survey (VVDS-Deep) spans the
apparent magnitude range $17.5<I_{AB}<24$ and contains about $8,700$ galaxies
over an area of $0.49~deg^2$. In addition to VIMOS spectroscopy, the field has
a multi-wavelength photometric coverage: $B,~V,~R,~I$ from the VIRMOS Deep
Imaging Survey \citep[VDIS, ][]{lefevre04}, $u^*,~g',~r',~i',~z'$ from the
Canada France Hawaii Telescope Legacy
Survey\footnote{http://terapix.iap.fr/cplt/table\_syn\_T0006.html}
\citep[CFHTLS-D1,][]{goranova09,coupon09}, and a partial coverage in $J$ and
$K$ bands from VDIS \citep{iovino05}.

Unless otherwise stated, in this analysis we use only the galaxies with secure
redshift, i.e. which have a redshift confidence level greater than $80\%$
\citep[flag 2 to 9, see][]{lefevre05a}. This provides us with a spectroscopic
sample of $6,582$ galaxies. The accuracy of the redshift measurements is of
$\sigma_z=9.2\cdot10^{-4}$. The observational spectroscopic strategy allows us
to reach an average spectroscopic sampling rate of $27\%$ across the
field. The observations, survey strategy, and basic properties of the
VVDS-Deep sample are described in detail in \citet{lefevre05a}. The VVDS-Deep
spectroscopic catalogue is publicly available through the CENter for COSmology
database (CENCOS) site\footnote{http://cencos.oamp.fr/}.

Absolute magnitudes for VVDS-Deep galaxies have been computed by
\citet{ilbert05} and properly k-corrected using the spectral template that
best fits the $B$, $V$, $R$, $I$ photometry. The VVDS-Deep apparent and
absolute magnitudes used in this study are not corrected for intrinsic dust
attenuation.

\begin{table*}
  \caption{Definition and properties of VVDS-Deep samples.}
  \centering
  \begin{tabular}{cccccc}
  \hline\hline
  \multicolumn{3}{l}{VVDS-Deep} \\
  \hline
  Redshift interval & Number of galaxies & $M^{mean}_B$ & $z^{mean}$ & $r_0$ &
  $\gamma$ \\
  \hline
  \multicolumn{6}{c}{All galaxies} \\
  $0.2 < z < 0.5$ & 1244 & -17.56 & 0.36 & $2.45^{+0.36}_{-0.42}$ & $1.68^{+0.09}_{-0.11}$ \\
  $0.5 < z < 0.7$ & 1430 & -18.60 & 0.61 & $2.63^{+0.20}_{-0.22}$ & $1.56^{+0.06}_{-0.06}$ \\
  $0.7 < z < 0.9$ & 1389 & -19.30 & 0.80 & $2.85^{+0.34}_{-0.36}$ & $1.59^{+0.05}_{-0.05}$ \\
  $0.9 < z < 1.1$ & 1020 & -19.84 & 0.99 & $2.71^{+0.24}_{-0.26}$ & $1.73^{+0.10}_{-0.09}$ \\
  $1.1 < z < 1.3$ &  562 & -20.39 & 1.18 & $3.17^{+0.12}_{-0.14}$ & $1.96^{+0.06}_{-0.05}$ \\
  $1.3 < z < 2.1$ &  434 & -21.05 & 1.48 & $3.51^{+0.32}_{-0.36}$ & $1.77^{+0.45}_{-0.04}$ \\
  \hline
  \multicolumn{6}{c}{Red galaxies} \\
  $0.2 < z < 0.7$ & 795 & -18.61 & 0.51 & $3.49^{+0.24}_{-0.34}$ & $1.87^{+0.16}_{-0.10}$ \\
  $0.7 < z < 1.1$ & 798 & -19.97 & 0.88 & $4.03^{+0.26}_{-0.26}$ & $1.62^{+0.04}_{-0.04}$ \\
%  $1.1 < z < 2.1$ & 219 & -21.05 & 1.27 & $4.01^{+0.50}_{-0.52}$ & $1.80^{+0.30}_{-0.30}$ \\
  \hline
  \multicolumn{6}{c}{Blue galaxies} \\
  $0.2 < z < 0.7$ & 1879 & -17.91 & 0.48 & $2.29^{+0.14}_{-0.14}$ & $1.61^{+0.04}_{-0.04}$ \\
  $0.7 < z < 1.1$ & 1611 & -19.32 & 0.88 & $2.45^{+0.06}_{-0.06}$ & $1.73^{+0.07}_{-0.07}$ \\
  $1.1 < z < 2.1$ &  777 & -20.58 & 1.32 & $3.21^{+0.14}_{-0.16}$ & $1.78^{+0.06}_{-0.05}$ \\
  \hline
  \label{table}
\end{tabular}
\end{table*}

\subsection{The galaxy formation model}

In this study, we take advantage of the semi-analytical model presented in
\citet{delucia07}, whose outputs are publicly available through the Millennium
Simulation database
site\footnote{http://www.mpa-garching.mpg.de/millennium/}. The model is
implemented on the Millennium Run N-body simulation, a large dark matter
N-body simulation which traces the hierarchical evolution of $2160^3$
particles between $z=127$ and $z=0$ in a cubic volume of
$500^3~h^{-3}~Mpc^3$. It assumes a $\Lambda CDM$ cosmological model with
$(\Omega_m=\Omega_{dm}+\Omega_b,~\Omega_\Lambda,~\Omega_b,~h,~n,~\sigma_8) =
(0.25,~0.75,~0.045, ~0.73,~1,~0.9)$. The resolution of the N-body simulation,
$8.6 \times 10^8~h^{-1} M_\odot$, coupled with the semi-analytical model
allows one to resolve haloes containing galaxies with a luminosity of $0.1L^*$
with a minimum of 100 particles. Haloes and sub-haloes are identified from the
spatial distribution of dark matter particles using a standard
friends-of-friends algorithm and the SUBFIND algorithm \citep{springel01}. All
sub-haloes are then linked together to construct the halo merging trees which
represent the basic input of the semi-analytical model. Details about the
simulation and the merger tree construction can be found in
\citet{springel05}.

Galaxies are simulated on top of the dark matter simulation using the
information contained in the halo merging trees. The semi-analytical model
(SAM) used in this study includes ingredients and methodologies originally
introduced by \citet{White91} and later refined by \citet{kauffmann00},
\citet{springel01}, \citet{delucia04}, \citet{croton06}, and
\citet{delucia07}. The model includes prescriptions for gas accretion and
cooling, star formation, feedback, galaxy mergers, formation of super-massive
black holes, and treatment of ``radio mode'' feedback from galaxies located at
the centres of groups or clusters of galaxies \citep[see][for
  details]{croton06}. One important aspect in the modelling of galaxy
properties at high redshift is dust extinction. This effect is particularly
critical when modelling galaxy magnitudes
\citep{kitzbichler07,fontanot09a}. In this work we have used the dust
extinction model of \citet{delucia07}, where dust extinction is parametrised
using a homogeneous interstellar medium component combined with a simple model
for molecular clouds attenuation around newly formed stars. In the following,
we will often refer to this model as the ``Munich'' model.

\subsection{Mock samples construction} \label{sec:mocks}

In order to compare the galaxy properties predicted by the model with
VVDS-Deep observations, we generated mock samples that cover the same volume
probed by the VVDS-Deep sample, as viewed by an observer at $z=0$. These have
been constructed using the MoMaF facility \citep{blaizot05}, which converts
the output of a galaxy formation model into simulated catalogues of
observations. In particular, the light-cone construction adopted in MoMaF
avoids replication and finite-volume effects in constructing mock survey
catalogues.

We built $100$ quasi-independent mock samples of $0.49~deg^2$ providing the
apparent magnitudes in the VDIS and CFHTLS filters for all galaxies. In
computing magnitudes, we use the appropriate filter transmission curves of the
VDIS and CFHTLS photometric bands.  These mock samples are linked to the main
SAM output, providing the full set of intrinsic galaxy properties that the
model calculates, as for instance galaxy absolute magnitudes and host halo
properties. The VVDS-Deep galaxy selection criterion, $17.5<I_{AB}<24$, has
been applied to the mock samples. These mock samples mimic the expected
VVDS-Deep sample with a spectroscopic sampling rate of $100\%$ and a uniform
sampling on the sky. Hereafter, we will refer to them as the ``complete'' mock
samples: \cmocks.

\begin{figure}
  \resizebox{\hsize}{!}{\includegraphics{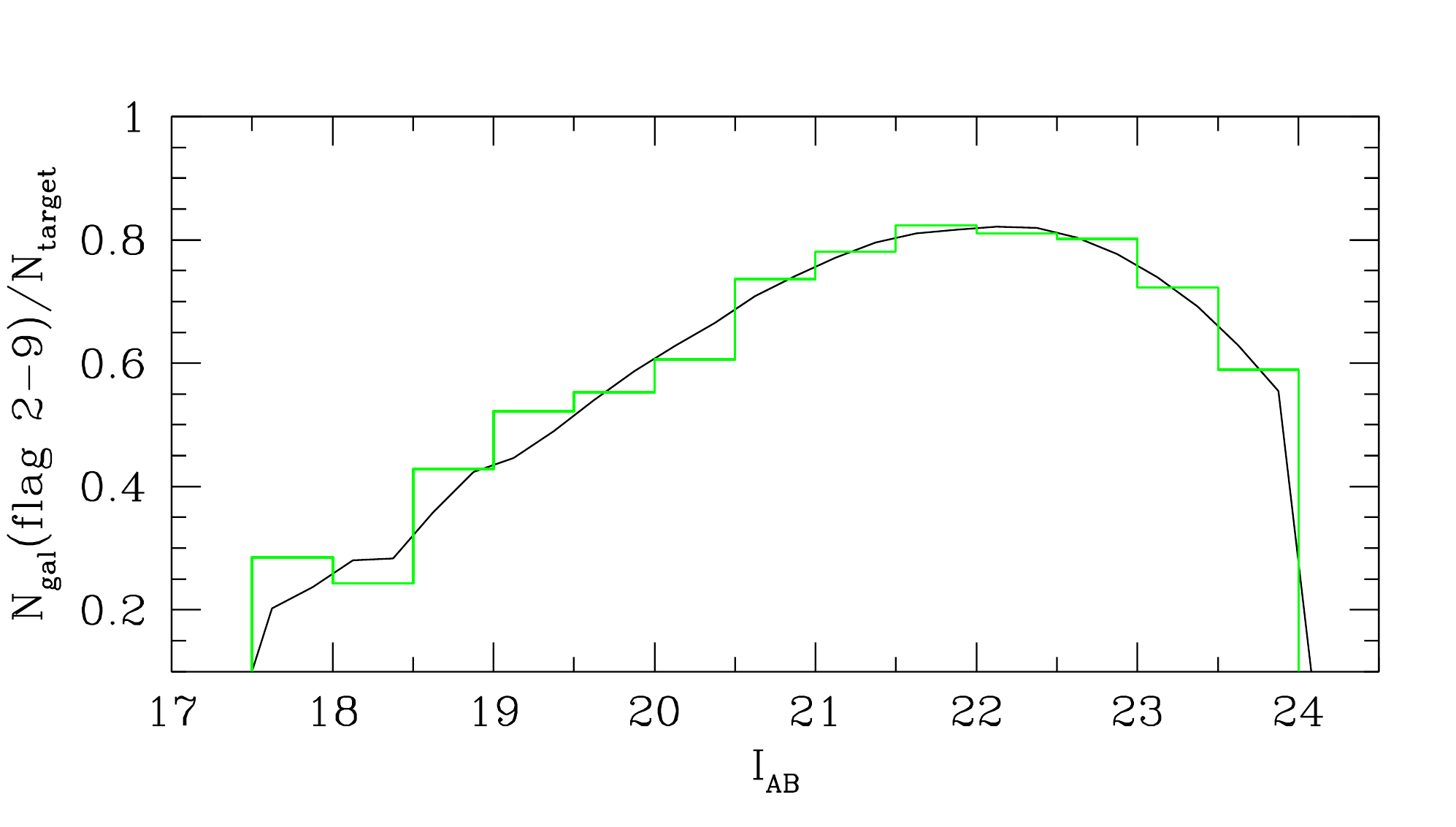}}
  \caption{Fraction of the number of galaxy with secure redshift (flag 2 to 9)
    over the total number of spectroscopic targets in the VVDS-Deep as a
    function of the $I$ apparent magnitude (histogram). The solid curve is the
    average fraction over the \omocks samples.}
  \label{ssr}
\end{figure}

\begin{figure}
  \resizebox{\hsize}{!}{\includegraphics{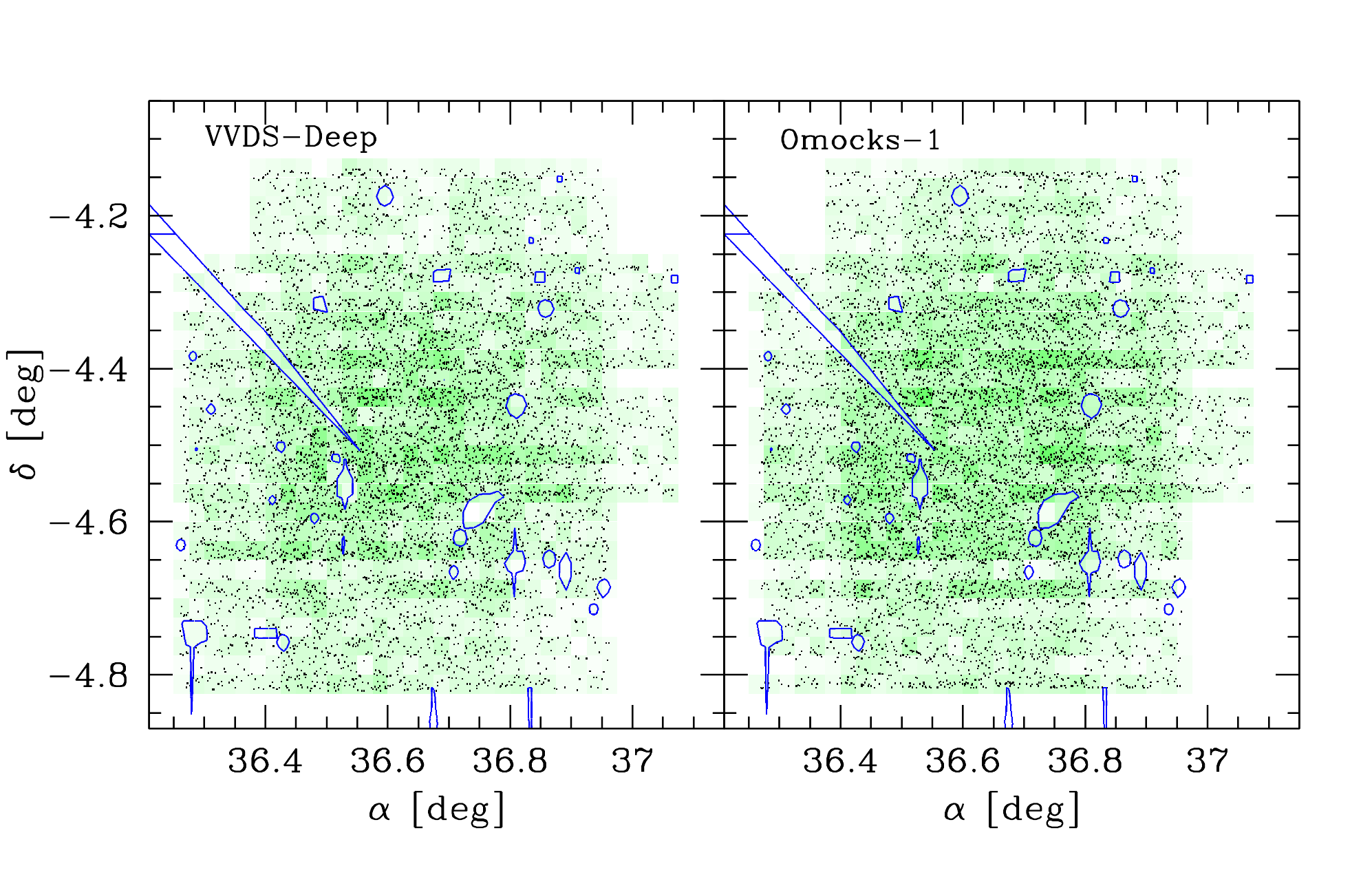}}
  \caption{Angular distribution of galaxies in the VVDS-Deep (left) and in the
    \omocks sample 1 (right). The galaxies are shown with the dots and the
    background coloured areas encode the surface density of objects. The solid
    contours encompass the regions discarded by the photometric mask.}
  \label{layout}
\end{figure}

From the \cmocks we constructed a second set of mock samples that include the
detailed observational selection function and biases of the VVDS-Deep sample,
and in particular, its complex angular sampling. These have been constructed
using the following procedure:
\begin{description}
\item[\bf Photometric mask.] In the VDIS images, objects falling in bad
  photometric regions or in areas where the presence of bright stars saturates
  the CCD, have been removed from the photometric catalogue. Consequently, no
  objects inside these regions have been targeted for spectroscopy. In
  practice, these regions have been identified and coded into a photometric
  mask, that we applied to the mock samples. This reproduces in the simulated
  catalogues the various empty regions present in the angular distribution of
  objects in the VVDS-Deep (see Fig. \ref{layout}).
\item[\bf Target sampling rate.] The SSPOC software \citep{bottini05} has been
  used to prepare VVDS-Deep observations, performing the design of the slit
  masks for the planned VIMOS pointings. This software, which accounts for the
  precise shape of the VIMOS field-of-view (4 quadrants delimited by an empty
  cross), enables one to optimise the slit positioning and to maximise the
  number of objects observed in spectroscopy from a given list of potential
  targets. Similarly, we applied SSPOC to the mock samples giving as input the
  list of VIMOS observed pointings in the VVDS-Deep. SSPOC requires the
  angular size of objects. Since this property is not attributed to mock
  galaxies by the model, we randomly assigned an apparent radius to mock
  galaxies, in a way to reproduce the observed distribution of apparent radii
  as a function of the selection magnitude in the VVDS-Deep. In the
  optimisation process, SSPOC tends to preferentially select objects with
  smaller apparent angular size. This introduces a mild bias against large
  galaxies, on average brighter, which we will refer to as the target sampling
  rate (TSR) in the following \citep{ilbert05}.
\item[\bf Spectroscopic success rate.] Galaxy redshifts in the VVDS-Deep
  sample have been determined from the observed spectra. In this process, a
  confidence class is given to each targeted object to quantify its type and
  the confidence level on the redshift determination \citep{lefevre05a}. In
  general, redshifts of galaxies with brighter apparent magnitudes, and in
  turn with higher signal-to-noise spectra, have a larger probability to be
  measured. The success in determining galaxy redshifts is thus a function of
  the apparent magnitude. This introduces a bias against faint objects which
  we will refer to as the spectroscopic sampling rate (SSR) in the following
  \citep{ilbert05}. To reproduce this selection bias in the mock samples, we
  randomly assigned a flag to each mock galaxy, in a way to match the
  dependence of the fraction of the different redshift confidence flags on the
  apparent magnitude in the VVDS-Deep. We kept in the mock samples the
  galaxies with flag 2 to 9, corresponding to objects with secure redshift in
  the VVDS-Deep \citep[confidence level greater than
    $80\%$,][]{lefevre05a}. Fig.~\ref{ssr} shows the fraction of galaxies with
  secure redshift measurements over the total number of spectroscopic targets
  in the VVDS-Deep, as a function of the apparent magnitude of selection. The
  solid curve in the same figure shows the mean measurement from the mock
  samples.
\end{description}

By applying this procedure to the \cmocks samples, we end up with a new set of
mock samples having on average a mean sampling rate of $27\%$ (identical to
that of the VVDS-Deep), and that include the detailed selection function and
observational biases of the VVDS-Deep spectroscopic sample. The angular
distribution of galaxies in the mock sample 1 obtained following this
procedure is shown as an example in Fig. \ref{layout} and compared to that of
the VVDS-Deep secure redshift sample. In the following, we will refer to these
mock samples as the ``observed'' mock samples: \omocks.

\begin{table*}
  \caption{Definition and mean properties of the Munich semi-analytical model
    \cmocks samples.}  \centering
  \begin{tabular}{cccccc}
  \hline\hline
  \multicolumn{3}{l}{SAM \cmocks} \\
  \hline
  Redshift interval & Mean number of galaxies & $M^{mean}_B$ & $z^{mean}$ & $r_0$ &
  $\gamma$ \\
  \hline
  \multicolumn{6}{c}{All galaxies} \\
  $0.2 < z < 0.5$ & 10662 & -17.34 & 0.37 & $3.37^{+0.44}_{-0.46}$ & $1.69^{+0.07}_{-0.06}$ \\
  $0.5 < z < 0.7$ &  8955 & -18.46 & 0.60 & $3.51^{+0.48}_{-0.50}$ & $1.68^{+0.08}_{-0.06}$ \\
  $0.7 < z < 0.9$ &  8246 & -19.10 & 0.80 & $3.47^{+0.36}_{-0.38}$ & $1.63^{+0.06}_{-0.05}$ \\
  $0.9 < z < 1.1$ &  5993 & -19.72 & 0.99 & $3.31^{+0.34}_{-0.36}$ & $1.58^{+0.05}_{-0.04}$ \\
  $1.1 < z < 1.3$ &  3583 & -20.30 & 1.19 & $3.31^{+0.24}_{-0.26}$ & $1.52^{+0.05}_{-0.04}$ \\
  $1.3 < z < 2.1$ &  4104 & -20.99 & 1.56 & $3.61^{+0.18}_{-0.20}$ & $1.52^{+0.03}_{-0.03}$ \\
  \hline
  \multicolumn{6}{c}{Red galaxies} \\
  $0.2 < z < 0.7$ & 5607 & -17.89 & 0.47 & $6.47^{+0.82}_{-0.94}$ & $1.94^{+0.09}_{-0.07}$ \\
  $0.7 < z < 1.1$ & 3087 & -19.53 & 0.87 & $6.31^{+0.86}_{-0.96}$ & $2.00^{+0.10}_{-0.07}$ \\
  $1.1 < z < 2.1$ &  725 & -20.58 & 1.23 & $5.91^{+1.12}_{-1.12}$ & $2.03^{+0.16}_{-0.14}$ \\
  \hline
  \multicolumn{6}{c}{Blue galaxies} \\
  $0.2 < z < 0.7$ & 14010 & -17.83 & 0.48 & $2.29^{+0.17}_{-0.18}$ & $1.44^{+0.05}_{-0.04}$ \\
  $0.7 < z < 1.1$ & 11152 & -19.31 & 0.88 & $2.35^{+0.17}_{-0.18}$ & $1.47^{+0.04}_{-0.04}$ \\
  $1.1 < z < 2.1$ &  6962 & -20.68 & 1.40 & $3.12^{+0.22}_{-0.23}$ & $1.44^{+0.04}_{-0.03}$ \\
  \hline
  \label{table2}
\end{tabular}
\end{table*}

\section{Number counts}

\subsection{Magnitude counts} \label{sec:magcounts}

We compare the galaxy magnitude counts in the VDIS $I$-band (selection band)
and in the CFHTLS $u^*, g', r', i', z'$ bands with those predicted by the
SAM. In this comparison, we use all galaxies with VDIS photometry for $I$-band
counts, while for $u^*, g', r', i', z'$ counts, we use all galaxies with
CFHTLS-D1 photometry falling in the VDIS field. The effective areas probed are
$1.19~deg^2$ and $0.89~deg^2$ respectively. We used \cmocks samples of same
angular sizes.  The VDIS and CFHTLS photometric catalogues reach limiting
magnitudes ($80\%$ completeness for point-like sources) of 24.6, 26.2, 25.9,
25.4, 25.1, 24.6 respectively for $I, u^*, g', r', i', z'$ bands
\citep[see][]{mccracken03,goranova09}. When computing the magnitude counts in
the VVDS data, we carefully excluded stars using colour-colour diagrams
following \citet{mccracken03}.

\begin{figure}
  \resizebox{\hsize}{!}{\includegraphics{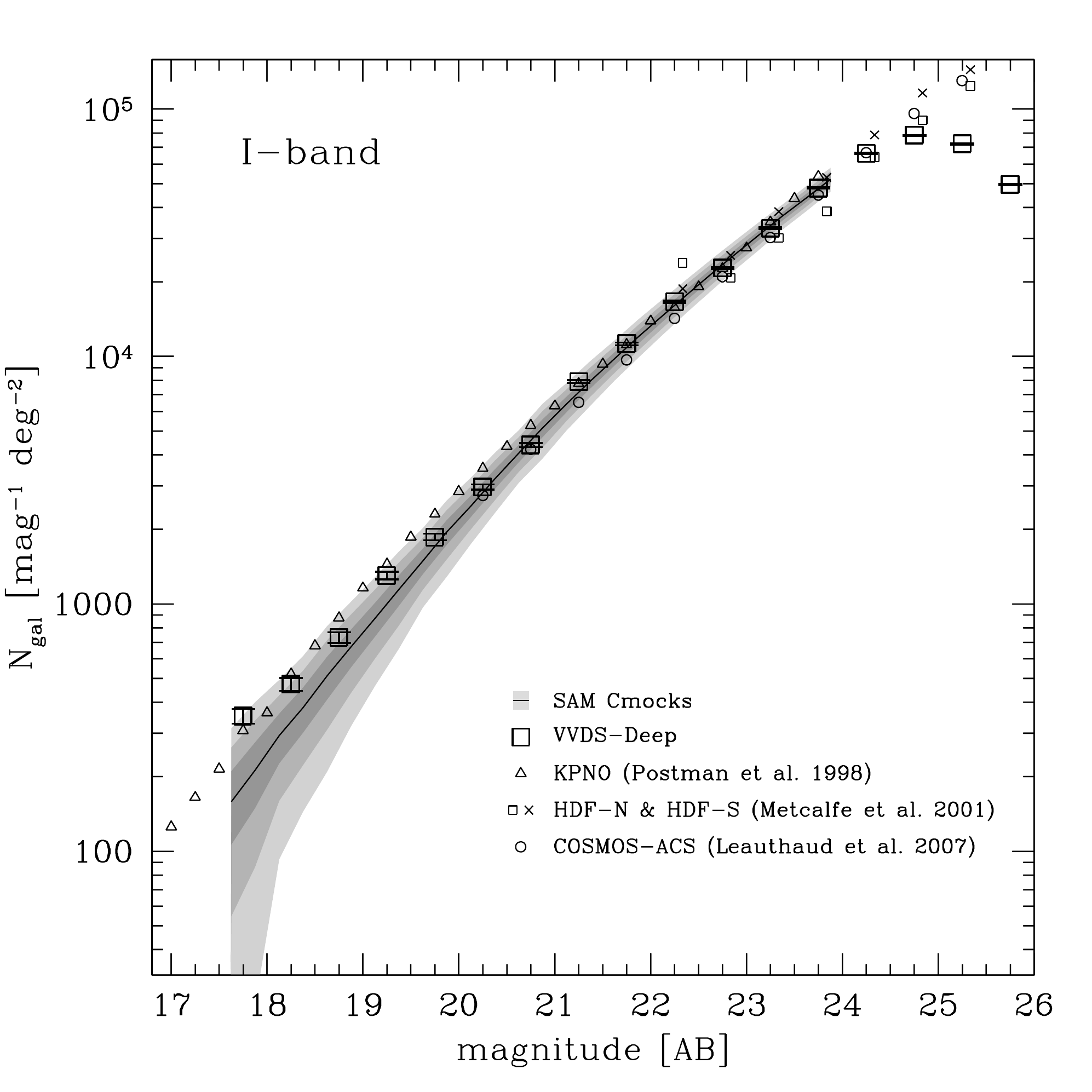}}
  \caption{Comparison of the magnitude counts in the VDIS $I$-band with SAM
    predictions and previously reported measurements by \citet{postman98} in
    the KPNO survey, \citet{leauthaud07} in the COSMOS-ACS field, and
    \citet{metcalfe01} in the Hubble Deep Field-North \& South. The VVDS-Deep
    counts are shown with the empty squares and the error bars correspond to
    their associated Poisson errors. The solid curve and associated shaded
    areas are respectively the mean and the 1$\sigma$, 2$\sigma$, and
    3$\sigma$ field-to-field dispersions among \cmocks samples.}
  \label{Nm_compi}
\end{figure}

\begin{figure*}
  \centering
  \includegraphics[width=12cm]{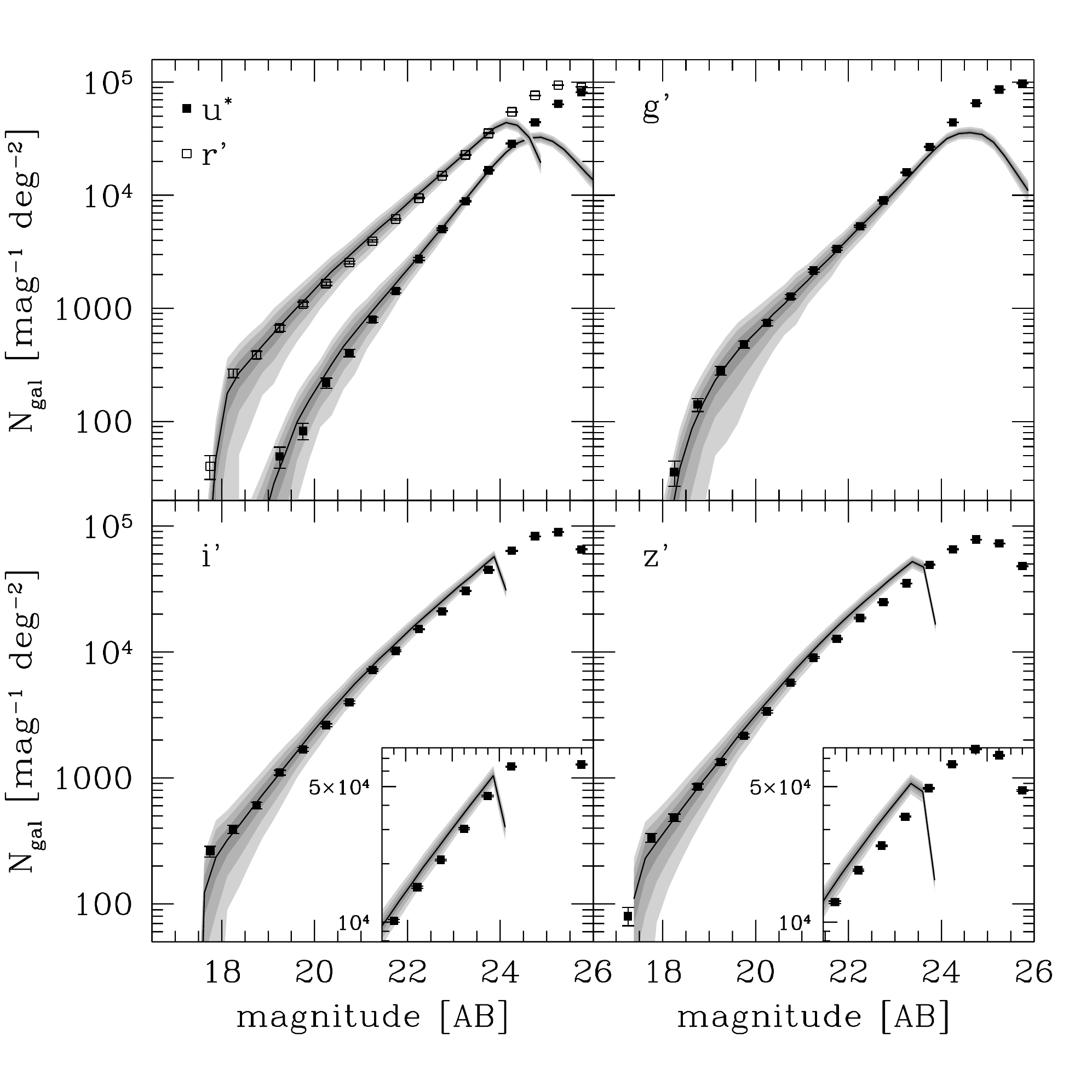}
  \caption{Comparison of the magnitude counts in the CFHTLS
    $u^*,~g',~r',~i',~z'$ optical bands with SAM predictions for $17.5<I<24$
    galaxies. The CFHTLS counts are shown with the squares and the error bars
    correspond to their Poisson errors. The curves and associated shaded areas
    are respectively the mean and the 1$\sigma$, 2$\sigma$, and 3$\sigma$
    field-to-field dispersions among the \cmocks samples. SAM predictions are
    for samples explicitly cut at $I=24$, while CFHTLS counts extend above
    this limit. All galaxies with $I<17.5$ have been removed from the counts.}
  \label{Nm_comp}
\end{figure*}

We first compare the $I$-band galaxy counts in the VDIS field with
published measurements from large and deep photometric surveys. This
comparison is shown in Fig. \ref{Nm_compi}. We find a very good agreement
between the VVDS and other published $I$-band counts in the literature
\citep[see also][]{mccracken03}. When comparing to the SAM predictions, we
find that the predicted $I$-band counts match very well the VVDS-Deep ones at
$I>20$.  The model, however, slightly underpredicts the number of galaxies
brighter than $I\simeq20$. On average, the mock samples contain $10.3\%$
less galaxies than observed at $17.5<I<24$ in the VVDS-Deep. In
Fig. \ref{Nm_comp} we compare the predicted magnitude counts in the CFHTLS
bands with the observational measurements. Note that in this figure, the
CFHTLS counts extend above $24~{\rm mag}$, while the SAM mock samples have
been cut at $I=24$ to match the VVDS-Deep selection function. We find that the
observed and predicted counts are in very good agreement, except in the $i'$-
and $z'$-bands where the SAM predicts more galaxies at the faintest
magnitudes. While in the $i'$-band the counts are compatible within the
3$\sigma$ field-to-field dispersion among the mock samples, the $z'$-band
counts deviate from the VVDS-Deep at more than 3$\sigma$ for magnitudes
fainter than $z'=22$.

Similar small discrepancies between model predictions and observed magnitude
counts were found by \citet{kitzbichler07} who, however, used observational
measurements based on smaller samples. In particular, they compared the counts
in the $B$, $R$, $I$ bands as observed over an area of $0.2~deg^2$ in the
Hubble Deep Field North (HDF-N), and in $K$-band, in fields of area smaller
than $0.1~deg^2$. They found a very good agreement between model predictions and
observations in the $B$, $R$, $I$ bands, with only a small underestimation of
the predicted counts at $I<19.5$.  For $K$-band counts, the agreement they
found was less good with the counts becoming discrepant at $K>21$ by a factor
of 2. These results are fully consistent with ours.

\subsection{Redshift distribution}

\begin{figure}
  \resizebox{\hsize}{!}{\includegraphics{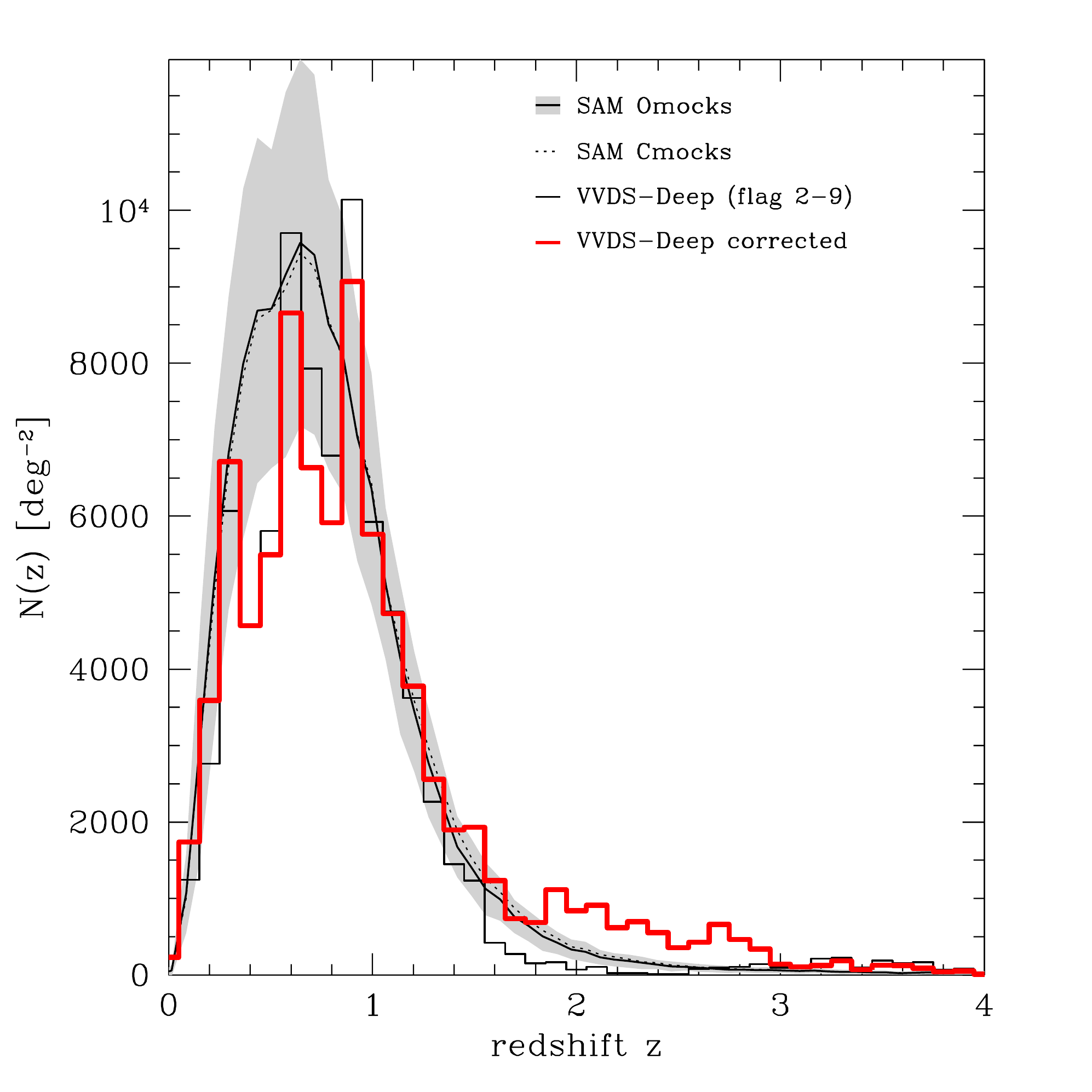}}
  \caption{Comparison of the VVDS-Deep with the SAM redshift distributions for
    $17.5<I_{AB}<24$ galaxies. The VVDS-Deep $N(z)$, while including only flag
    2 to 9 galaxies, is shown with the solid thin histogram. The thick
    histogram corresponds to the corrected N(z) in the VVDS-Deep when
    accounting for the observational biases of the survey (see text for
    details). The solid curve and associated shaded area are the mean and the
    1$\sigma$ field-to-field dispersion among the \omocks, while the dotted
    curve is the mean prediction of the \cmocks.}
  \label{Nz_comp}
\end{figure}

To provide a fair comparison of the VVDS-Deep redshift distribution with that
predicted by the SAM, we first consider only the VVDS-Deep galaxies with
secure redshifts (flag 2 to 9), which represent $75\%$ of the full galaxy
sample. The remaining $25\%$ of the sample is made of galaxies with a redshift
confidence level of about $50\%$ (flag 1, $17\%$) and unclassified objects for
which no redshift measurement has been possible (flag 0, $8\%$). Note that
this latter class possibly includes stars. As redshifts are available only for
about $27\%$ of the galaxies with $17.5<I<24$, we normalise the redshift
distribution to the total number of galaxies in the parent photometric
catalogue. The corresponding redshift distribution is shown in
Fig. \ref{Nz_comp} with the solid thin histogram. This normalisation assumes
that the spectroscopic sample is statistically representative of the parent
photometric catalogue and that there are no biases introduced by the
particular spectroscopic targeting strategy adopted. In reality, both these
assumptions are incorrect.

In order to account for the observational biases and estimate their influence
on the observed redshift distribution, we use the information of the survey
target sampling rate (TSR) and spectroscopic sampling rate (SSR) available for
each galaxy of the sample. These two functions have been originally defined to
correct for the observational biases in the measurement of the galaxy
luminosity function \citep{ilbert05,zucca06}. The SSR accounts for the
fraction of objects without a reliable redshift determination (flag 0 and flag
1 galaxies) and corrects for the fact that the success rate of redshift
measurements decreases at fainter apparent magnitudes. The TSR instead,
corrects for the selection biases introduced by SSPOC software in the
spectroscopic mask preparation \citep{bottini05}, e.g. its tendency to target
objects with small apparent angular size. Within the Ultra-Deep part of the
VVDS survey \citep{lefevre10}, a small fraction of flag 0, flag 1, and flag 2
galaxies have been reobserved. These reobservations, representing $4\%$ of the
total number of objects in the spectroscopic catalogue, have permitted us to
refine the measurement of the TSR and SSR functions, allowing us to better
statistically account for the spectroscopic incompleteness. The
VVDS-Ultra-Deep observations, as well as details on the calculation of the TSR
and SSR are given in \citet{lefevre10}.

We correct the raw redshift counts, including flag 1 galaxies, by weighting
each galaxy by $w=(\mbox{TSR} \times \mbox{SSR})^{-1}$. The corrected redshift
distribution is shown in Fig. \ref{Nz_comp} with the solid thick
histogram. The solid curve and associated shaded area correspond to the
predicted mean $N(z)$ and the $1\sigma$ field-to-field dispersion among
\omocks samples, which provides an estimate of the sample variance in the
model. The predicted mean $N(z)$ of the \cmocks samples is plotted with the
dotted curve.

We first note that VVDS-Deep observational biases have little effect on the
shape of the predicted redshift distribution in the mock samples: the mean
$N(z)$ obtained from \omocks is very similar to that of \cmocks. The SAM well
reproduces the shape of the observed distribution between $z\simeq 1$ and
$z\simeq 1.8$ while it shows an excess of about 14 per cent with respect to
the observations at $0.2<z<1$. Similar trends were found in
\citet{kitzbichler07} who noted that the predicted redshift distribution was
higher than observed over the redshift range $0.5 < z < 1.5$ for $K<21.8$
galaxies. As we will discuss in the next section, this excess is related to an
excess of red galaxies at $z<1$ in the model. Furthermore, we find that the
model does not account for the tail of the distribution at redshifts above
$z\simeq2$ and that extends to $z\simeq4$ in the VVDS-Deep sample. In fact the
SAM does not predict any galaxy at $z>3$ with $I<24$. The dip observed at $1.8
< z < 3$ in the VVDS-Deep uncorrected N(z) is a purely observational effect
usually referred as the ``redshift desert''. It is due to the lack of spectral
features in galaxy spectra observed within the wavelength window function of
the spectrograph at these redshifts \citep{lefevre05a}. VVDS-Ultra-Deep
reobservations, based on spectra measured on a larger wavelength range (VIMOS
LR-Blue plus LR-Red grisms), allow us to correct for this observational effect
and to repopulate the ``redshift desert'' at $1.8<z<3$. Indeed, a large
fraction of the reobserved galaxies falls in this part of the redshift
distribution, and thus by properly weighting the original flag 1 to 9 galaxies
we are able to statistically account for the missing fraction of objects at
these redshifts (see Le F\`evre et al., 2010).

\subsection{Galaxy luminosity and intrinsic colour distributions}

\begin{figure}
  \resizebox{\hsize}{!}{\includegraphics{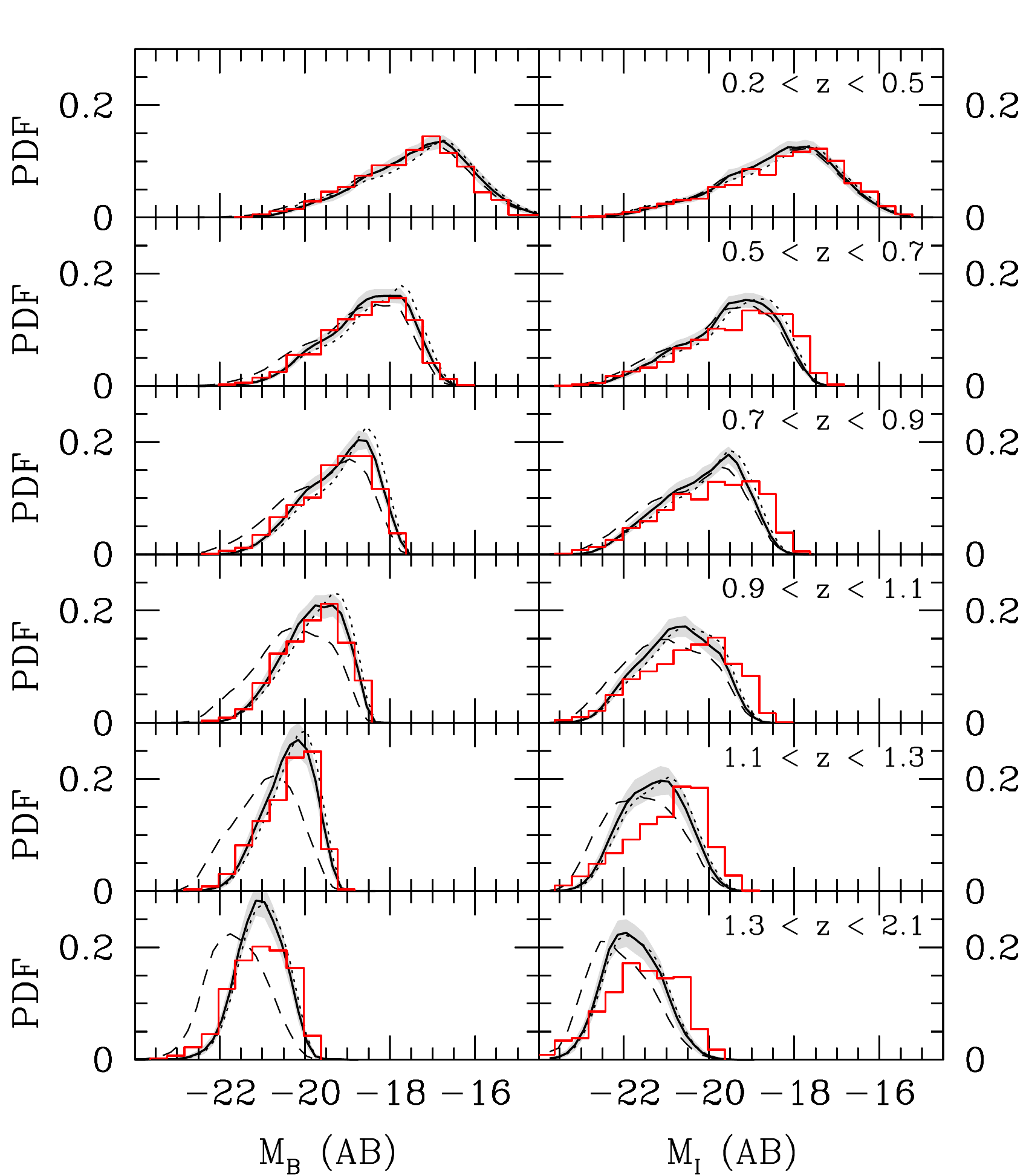}}
  \caption{Comparison of the rest-frame $B$-band (right panels) and $I$-band
    (left panels) distributions in six magnitude-limited samples from $z=0.2$
    to $z=2.1$. The histograms are VVDS-Deep measurements while the curves and
    associated shaded areas correspond to the mean and the 1$\sigma$
    field-to-field dispersion among the \omocks samples. The dashed curves are
    the mean distribution in the \omocks without including dust extinction in
    the model, while the dotted ones are the mean predictions of the \cmocks.}
  \label{mabsct}
\end{figure}

\begin{figure}
  \resizebox{\hsize}{!}{\includegraphics{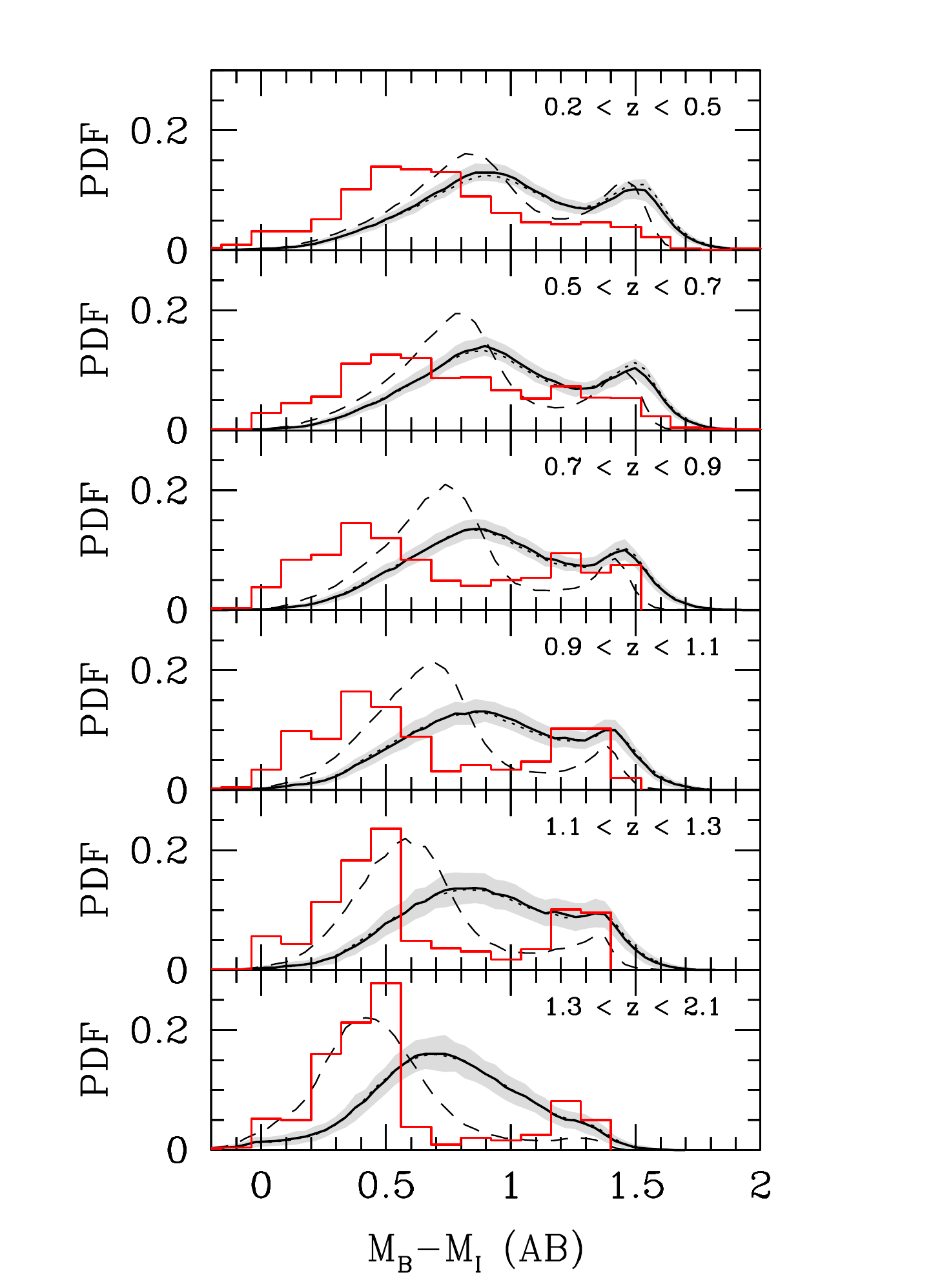}}
  \caption{Comparison of the rest-frame $B-I$ distribution in six
    magnitude-limited samples from $z=0.2$ to $z=2.1$. In each panel, the
    histogram corresponds to the VVDS-Deep measurement while the curve and
    associated shaded area correspond to the mean and the 1$\sigma$
    field-to-field dispersion among the \omocks samples. The dashed curves are
    the mean distribution in the \omocks without including dust extinction in
    the model, while the dotted ones are the mean predictions of the \cmocks.}
  \label{bimod}
\end{figure}

Galaxies are found to have a bimodal colour distribution and rest-frame
colours are commonly used to differentiate red massive early-type from blue
star-forming late-type galaxies. Here we define these two populations on the
basis of the rest-frame $B-I$ colour distribution. We made this particular
choice because it corresponds to the optimal rest-frame colour that can be
measured both in the VVDS-Deep and in the model (the rest-frame $U$-band is
not available in the SAM). We considered six redshift intervals from $z=0.2$
to $z=2.1$, and compare the rest-frame $B$-band, $I$-band, and $B-I$
distributions from the model with the observational measurements. The
definition of the samples and their basic properties are given in Table
\ref{table} and Table \ref{table2}. The galaxy luminosity and rest-frame
colour distributions (normalised to unity) in the VVDS-Deep and the SAM are
presented in Fig. \ref{mabsct} and Fig. \ref{bimod}.

The predicted rest-frame $B$-band distributions in the different redshift
intervals are in good agreement with VVDS-Deep observations (the apparent
discrepancy in the peak height in the last high-redshift interval is not
statistically significant given the small number of galaxies involved). This
indicates that the $B$-band luminosity function may be well reproduced in the
SAM up to $z\simeq2$ as also found by \citet{kitzbichler07} at
$0.2<z<1.2$. Instead, we find that the rest-frame $I$-band distributions are
systematically skewed towards bright magnitudes in the SAM. The effect is
particularly important at $z>0.9$ where the model tends to significantly
overestimate the number of galaxies with $-23<M_I<-21$. A similar trend was
found in \citet{kitzbichler07} when confronting the $K$-band luminosity
function, although the small observational samples they used did not allow
them to reach firm conclusions.

When comparing the intrinsic colour distributions, we find that the rest-frame
$B-I$ colour distribution is clearly bimodal both in the VVDS-Deep and in the
model, but the agreement is only qualitative. As shown in Fig. \ref{bimod},
the SAM does not reproduce quantitatively the observed intrinsic colour
distributions: there are many fewer very blue galaxies (i.e. with
$B-I\simeq0.3$) and much more ``green valley'' galaxies (i.e. with
$B-I\simeq1$) in the model than in the observations, at all probed
redshifts. In addition, the model predicts an excess of red galaxies at low
redshift. It could be argued that part of the discrepancies between the SAM
and observed colours could be related to uncertainties in the modelling of
dust extinction \citep[e.g.][]{kitzbichler07,fontanot09a}.  The dashed line in
Fig. \ref{bimod} shows the rest-frame $B-I$ colour distribution in the SAM
without including dust extinction in the model galaxies. In that case, when
comparing to VVDS-Deep colour distributions, one finds that while in the
highest-redshift intervals the predicted and observed colour distributions are
quite similar, the lack of blue galaxies in the model is remarkable at
$z<1.1$. \citet{franzetti07} show that the fixed apparent magnitude selection
of the VVDS-Deep sample can in principle introduce a mild bias in the
intrinsic colours, partially displacing rest-frame $U-V$ colours towards the
blue at $z>1.2$. However, they find that this effect is marginal and cannot be
invoked to explain the discrepancies found in the SAM at all probed
redshift. Intrinsically, the SAM may not form enough very blue galaxies.

\begin{figure}
  \resizebox{\hsize}{!}{\includegraphics{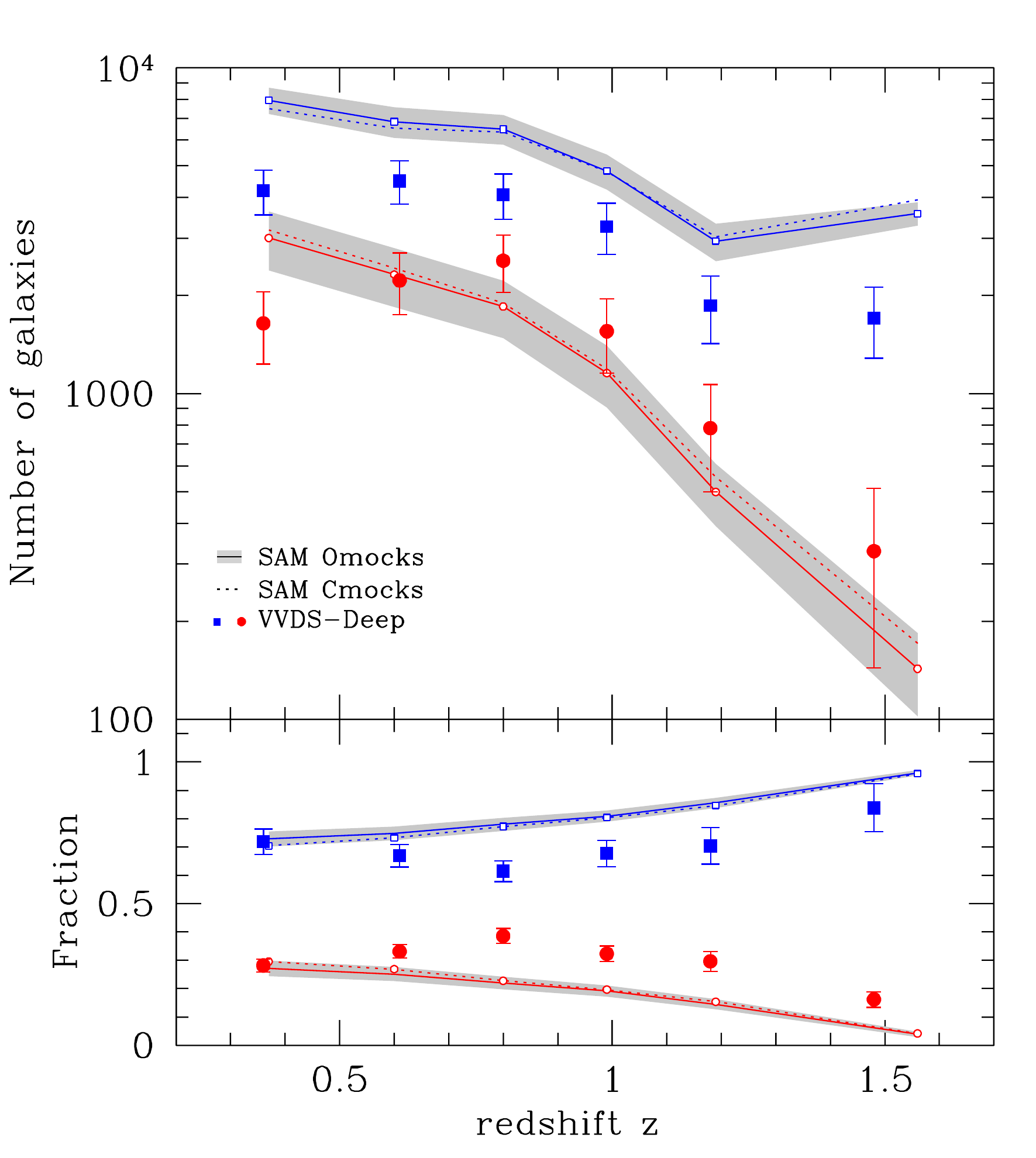}}
  \caption{Total number and fraction of red (filled circles) and blue (filled
    squares) galaxies as a function of redshift in the VVDS-Deep and in the
    SAM. The symbols with error bars are the VVDS-Deep measurements while
    solid curves and associated shaded areas correspond to the predicted mean
    fractions and 1$\sigma$ field-to-field dispersions among the \omocks. The
    dotted curves are the mean predictions of the \cmocks. In both upper and
    lower panel, higher curves and symbols correspond to blue galaxies.}
  \label{fraction}
\end{figure}

Because of the discrepancies between the predicted and observed colour
distributions, it is difficult and possibly meaningless to define ``blue'' and
``red'' galaxies using the same colour cut. We therefore opted to use a
different colour cut for the SAM and the VVDS-Deep sample, with the aim of
separating blue and red populations on the basis of the colour bimodality. In
the VVDS-Deep we use a cut at $(B-I)^{cut}=0.95$. \citet{zucca06} show that
galaxies selected above and below this value largely overlap with those
classified as early- and late-type galaxies using a more refined method based
on spectral energy distribution fitting. We adopt a larger value of
$(B-I)^{cut}=1.3$ to separate red and blue populations in the SAM. We compare
in Fig. \ref{fraction} the total number and the fraction of red and blue
galaxies at different redshifts.

We find a large difference in the number density of blue galaxies as predicted
by the SAM and observed in the VVDS-Deep. While the trends with redshift are
rather similar, the SAM predicts 50-80\% more blue galaxies than observed over
the whole redshift interval $0.2<z<1.6$ in the VVDS-Deep. In the observations
we find the presence of an already significant number of red galaxies at
$z\simeq1.5$. This number increases until $z\simeq0.8$ and then slightly
decreases with cosmic time. In contrast, the SAM predicts a monotonic increase
with time of the number of red galaxies. Above $z\simeq0.8$, the total number
of red galaxies is smaller in the model than in the observations but the trend
reverses at later epochs, where the total number of red VVDS-Deep galaxies starts
to decline and that of red model galaxies continues to rise slowly. These same
problems are evident when looking at the fractions of the two populations
(bottom panel in Fig. \ref{fraction}). The SAM predicts a monotonic decrease
(increase) of the fraction of blue (red) galaxies with time at variance with
VVDS-Deep sample, in which this trend reverses at $z<0.8$. Note that
we discuss the variation with cosmic time of the number and fraction of red
and blue galaxies in terms of \emph{apparent} increase or decrease in a given
apparent magnitude range, which does not necessarily imply \emph{real}
increase or decrease in number density at these redshifts.

At redshifts higher than $z\simeq0.8$, the number of red galaxies is slightly
lower in the SAM with respect to observations, while blue galaxies appear to
be significantly more abundant at all cosmic epochs. This indicates that the
overabundance of galaxies previously seen in the redshift distribution below
$z\simeq 0.8-1$ in the SAM (Fig. \ref{Nz_comp}), is due to the presence of a
larger number of both blue (true at all redshifts for this colour) and red
galaxies.

The difference in the variation with cosmic time of the fraction of red and
blue galaxies and of the shape of the rest-frame $B-I$ colour distribution in
the SAM, suggests that these populations may have different histories of
formation and evolution than in the VVDS-Deep. In particular, SAM red galaxies
may start to form at later epochs and be forming continuously and more
efficiently up to present day, in contrast with what appears to be happen to
VVDS-Deep galaxies.

\section{Galaxy clustering}

The clustering properties of galaxies provide strong constraints on galaxy
formation models as they encode important information on how galaxies populate
dark matter haloes. We compare in this section the galaxy clustering as
inferred from the two-point correlation function in the SAM with VVDS-Deep
measurements, for the global population of $17.5<I<24$ galaxies.

\subsection{Two-point correlation function estimation} \label{sec:meth}

We estimate the real-space galaxy clustering using the standard projected
two-point correlation function, $w_p(r_p)$, that corrects for redshift-space
distortions due to galaxy peculiar motions. This is obtained by splitting the
galaxy separation vector into two components, $r_p$ and $\pi$, perpendicular
and parallel to the line of sight respectively \citep{peebles80,fisher94}, and
projecting the two-dimensional two-point correlation function $\xi(r_p,\pi)$
along the line-of-sight:
\begin{equation}
  w_p(r_p)=2\int_{0}^{\pi_{max}}\xi(r_p,\pi)d\pi \,\,\,\, . \label{eq:wprp}
\end{equation}
We use the standard \citet{landy93} estimator to compute $\xi(r_p,\pi)$. In
practice, to obtain \wprp, we integrate $\xi(r_p,\pi)$ up to
$\pi_{max}=20~h^{-1}~Mpc$. We adopt this value because we find that, given the
volume of the survey, this value is large enough so as to minimise the noise
introduced at large $\pi$ by the uncorrelated pairs in the data
\citep[][]{pollo05}. Errors in the VVDS-Deep are estimated through the
blockwise bootstrap resampling technique \citep[e.g.][]{porciani02}, which
allows us to account for sample variance in the field and provides fair error
estimates, very similar to those obtained using the Jackknife resampling
technique \citep{norberg09}. Since the transverse dimension of the survey is
small, to generate the different resamplings we divide each sample in slices
along the radial direction \cite[e.g.][]{delatorre09,meneux09}. For each of
our samples, we use $300$ resamplings by bootstrapping $6$ slices of equal
volume.  We estimated the errors in the SAM by computing the field-to-field
variance among the 100 mock samples. We explicitly verified that the bootstrap
error estimates agree with the ensemble errors obtained from the mock samples.

The VVDS-Deep sample has a complex angular sampling as shown in
Fig. \ref{layout}. To measure the projected correlation function we follow the
method originally introduced by \citet{pollo05} as improved by
\citet{delatorre09}. This method allows us in particular to correct \wprp
measurements for the inhomogeneous sampling on the sky and the incompleteness
on small angular scales due to slit mask design.  The angular inhomogeneous
sampling, i.e. the fact that the sampling rate varies with the angular
position, is accounted for by reproducing in the random sample, used to
estimate $\xi(r_p,\pi)$, the same variations of sampling. This entails
accounting for the precise shape of the VIMOS field-of-view and the
coordinates of the observed pointings. A similar technique was applied in the
previous analysis of the VVDS-Deep
\citep{pollo06,meneux06,delatorre07,meneux08}. The improved method used here
includes a more accurate pair weighting scheme to correct for missed angular
pairs. Indeed, we correct for the incompleteness on small angular scales by
weighting each galaxy-galaxy pair by the ratio of the number of pairs in the
spectroscopic sample to that in the parent photometric catalogue (free from
angular incompleteness), as a function of the angular separation. We refer the
reader to \citet{pollo05} and \citet{delatorre09} for the full description of
the method used to account for the survey selection function and observational
biases in the measurement of \wprp.

\begin{figure}
  \resizebox{\hsize}{!}{\includegraphics{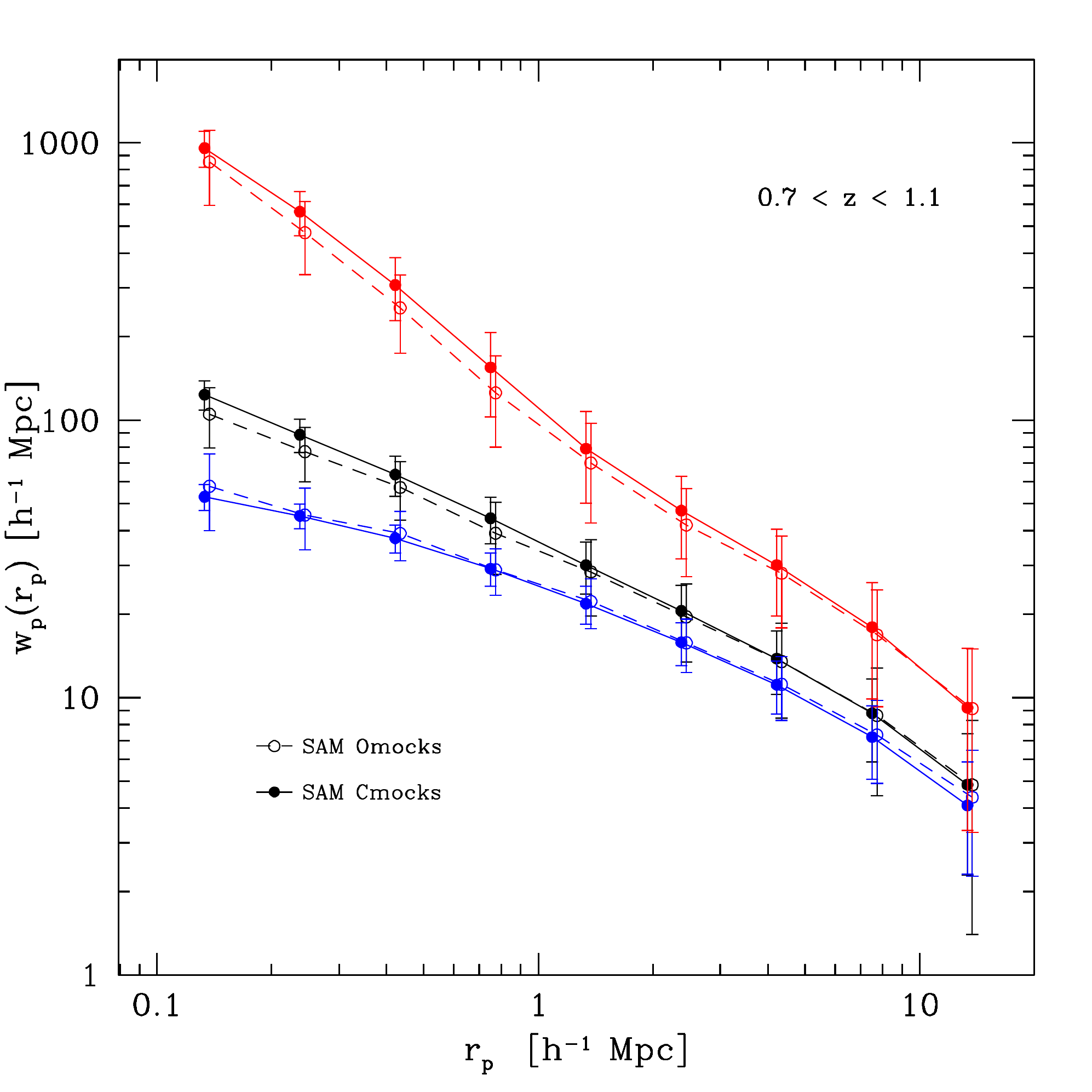}}
  \caption{Mean projected correlation function at $0.7<z<1.1$ for red (top
    curves), all (middle curves), and blue (bottom curves) galaxies in the
    \omocks (dashed curves) and \cmocks (solid curves) samples. The error bars
    correspond to the 1$\sigma$ field-to-field dispersion among the mock
    samples. \omocks points have been slightly displaced along $r_p$-axis to
    improve the clarity of the figure.}
  \label{ovsfmocks}
\end{figure}

In Fig. \ref{ovsfmocks} we present the estimated galaxy projected correlation
function at $0.7<z<1.1$ in the mock samples when including (\omocks) or not
(\cmocks) the detailed VVDS-Deep selection function for all, red, and blue
galaxies. We find that the method does not introduces any systematic errors on
the amplitude and shape of \wprp up to $r_p=15$\hmpc for the blue galaxy
population. For more strongly clustered population as incarnated by all and
red galaxies, our method tends to underestimate the amplitude of \wprp on
scales smaller than $\sim2-3$\hmpc by at maximum 15\%. In fact, although the
method is quite robust, we cannot entirely recover the small-scale clustering
information as we only sample, on average, 27\% of the galaxies in
spectroscopy. However, these systematic errors are smaller than the
statistical $1\sigma$ errors of the mock samples. In the following, we will
then use the \omocks and the VVDS-Deep secure redshift sample to perform
clustering comparisons.

While in general the shape of the observed correlation function deviates from
a pure power-law \citep[e.g.][]{zehavi04}, this simple parametrisation allows
one to quantify the clustering properties of galaxy samples and to easily
compare them. From the real-space two-point correlation function $\xi(r)$, the
correlation length $r_0$, which characterises the clustering strength, and the
slope $\gamma$ are obtained by fitting $\xi(r)$ to a power-law such as
$\xi(r)=(r/r_0)^{-\gamma}$. In the case of the projected (real-space)
correlation function $w_p(r_p)$, the power-law form transforms to
\citep{peebles80}, \begin{equation}
  w_p(r_p)=r_p\left(\frac{r_p}{r_0}\right)^{-\gamma}\frac
  {\Gamma\left(\frac{1}{2}\right)\Gamma\left(\frac{\gamma-1}{2}\right)}
  {\Gamma\left(\frac{\gamma}{2}\right)}~~, \end{equation} where $\Gamma$ is
the Euler Gamma function. In the present analysis we fit the \wprp
measurements on the range $0.1~{\rm h^{-1}~Mpc}<r_p<10~{\rm h^{-1}~Mpc}$ using the
generalised $\chi^2$ method. We use the full covariance matrix estimated from
the measurements to account for the correlations between the different $r_p$
bins in \wprp \citep[e.g.][]{pollo05}.

\subsection{Clustering of the global population} \label{sec:wpall}

\begin{figure*}
  \centering \includegraphics[width=12cm]{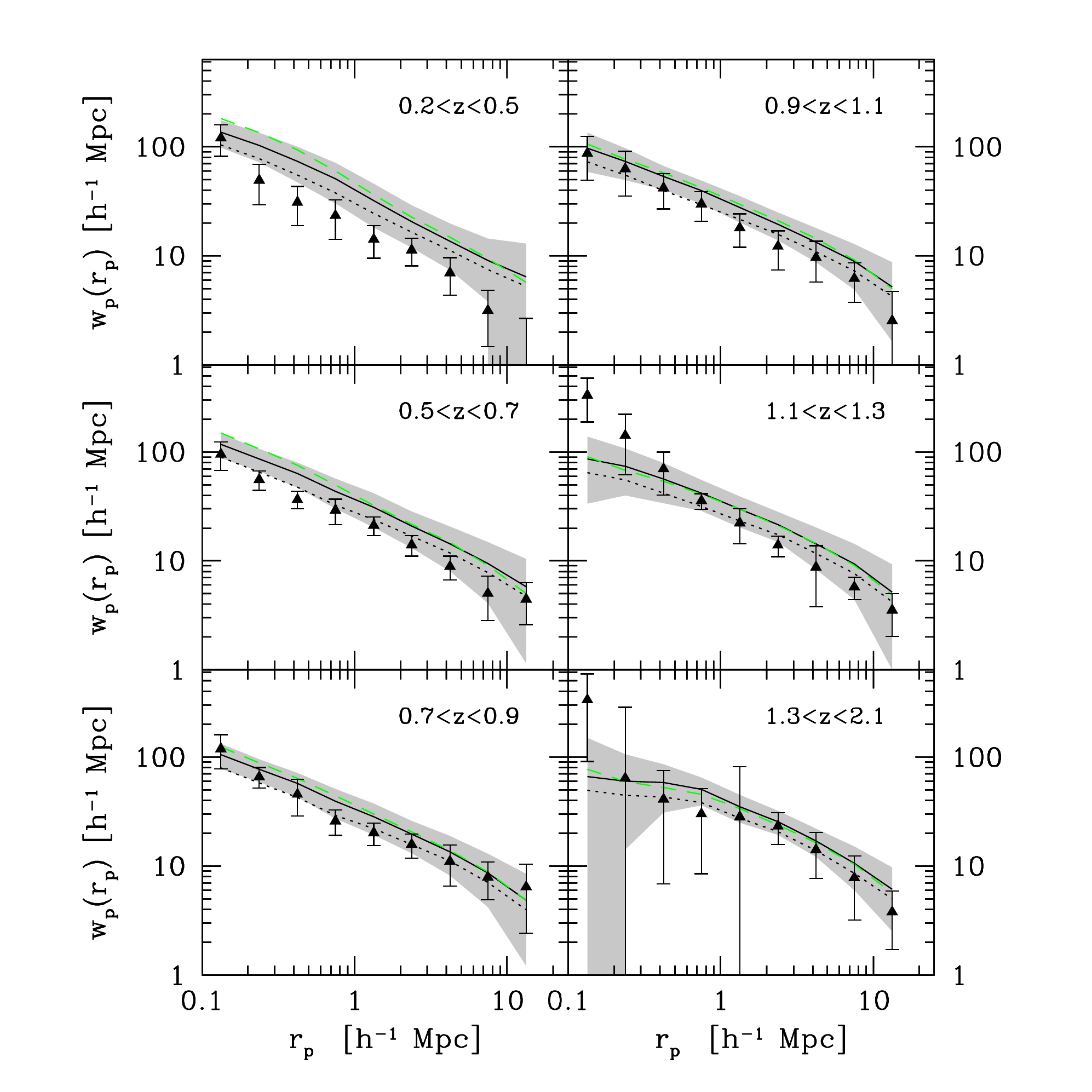}
  \caption{Comparison of the projected correlation functions as a function of
    redshift. In each panel, the filled triangles with error bars correspond
    to VVDS-Deep measurements while the solid curve and associated shaded area
    correspond to the mean and the 1$\sigma$ field-to-field dispersion among
    the \omocks samples. The dashed curves are the mean \cmocks predictions,
    while the dotted ones are the mean \wprp obtained by rescaling the
    predicted correlations function in the \omocks to $\sigma_8=0.81$ as
    described in Sec. \ref{sec:wpall}.}
  \label{wp_comp}
\end{figure*}

\begin{figure}
  \resizebox{\hsize}{!}{\includegraphics{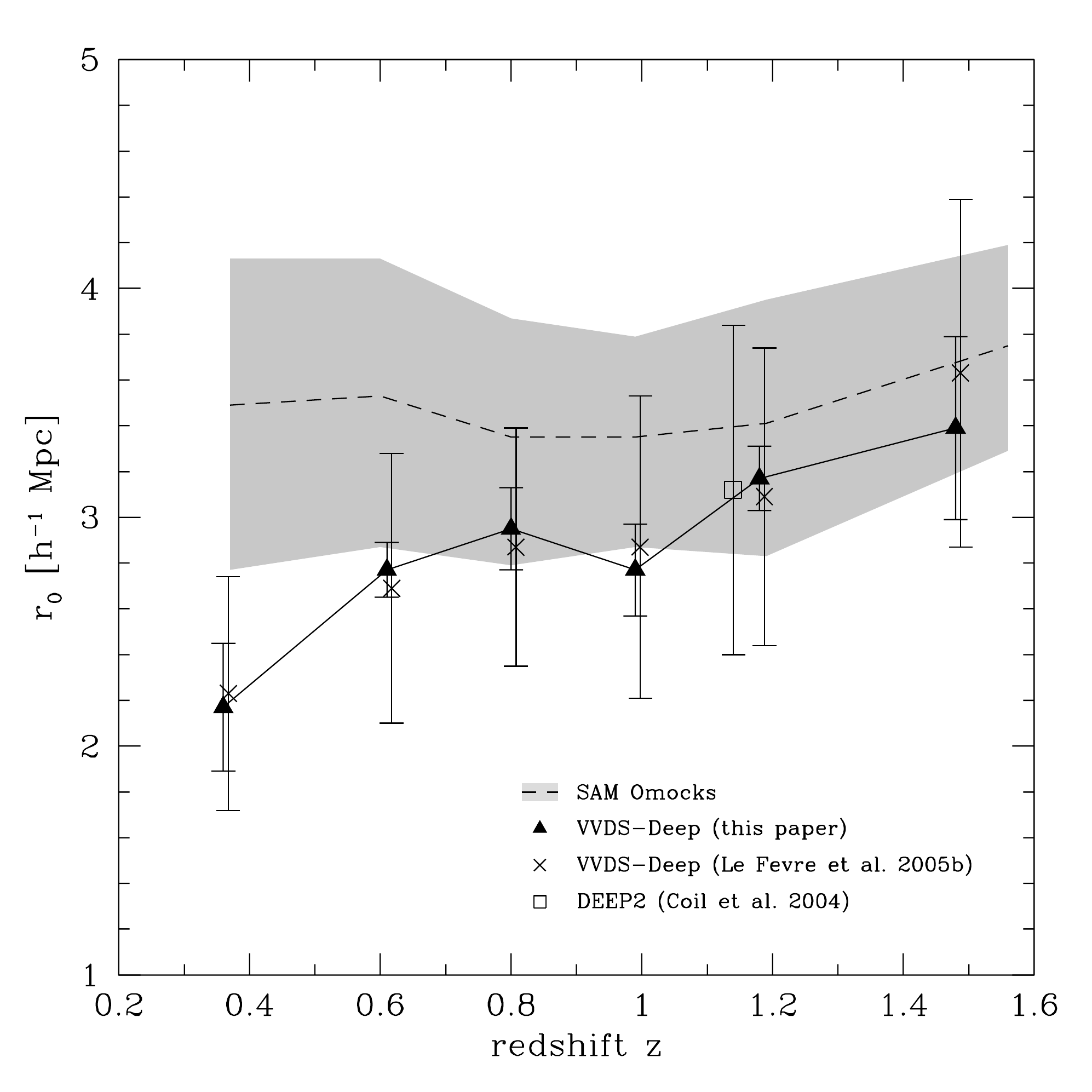}}
  \caption{Comparison of the correlation lengths as a function of
    redshift. The filled triangles with error bars are VVDS-Deep measurements
    while the dashed curve and associated shaded area correspond to the
    correlation length and error from the \omocks. We also report in this
    figure the previous VVDS-Deep measurements by \citet{lefevre05b} (crosses)
    and the one obtained by \citet{coil04} in the DEEP2 survey (open square).}
  \label{ro_comp}
\end{figure}

The clustering evolution of the global population of galaxies in the VVDS-Deep
has been measured by \citet{lefevre05b} in the six redshift intervals from
$z=0.2$ to $z=2.1$ previously defined. We update their measurements using the
more accurate method described in the previous section. The \wprp measurements
are shown in Fig. \ref{wp_comp} along with the mean \wprp among the SAM
\omocks samples.  We fit the \wprp with power laws and provide the best-fitted
parameters $r_0$ and $\gamma$ in Tab. \ref{table} and Tab. \ref{table2} for
both the VVDS-Deep and the SAM. We compare the measured correlation lengths in
Fig. \ref{ro_comp}. This figure shows that our measurements agree very well
within the uncertainties with those previously obtained by \citet{lefevre05b}
for the same field, as well as with results from the DEEP2 survey
\citep{coil04}.

We recall that for all clustering comparisons, we use model predictions
computed from the \omocks using the same method adopted for the VVDS-Deep
data. The Fig. \ref{wp_comp} shows that, when comparing \omocks (solid curves)
and \cmocks (dashed curves) measurements, one find a non-negligible bias in
the estimation of \wprp on small scales at $z<0.7$. As already pointed out by
\citet{lefevre05b}, here the incompleteness effects are enhanced by the small
volume and by the small number of objects in the observed samples at these
redshifts. This emphasises the importance (and the necessity) to include
detailed observational selection functions and biases in the mock samples in
order to carry out a fair comparison between observational measurements and
model predictions.

At $z<1.1$, the SAM predicts on average a higher clustering amplitude than
measured in the VVDS-Deep, while the slopes of the predicted and observed
correlation functions are similar. In addition, while the correlation length
in the VVDS-Deep increases slightly with increasing redshift, the model
predicts a roughly constant correlation length over all the redshift range
probed by the observations. Some evolution of the overall clustering is
expected, because by selecting galaxies at increasing redshift in a
magnitude-limited sample, we probe intrinsically more luminous galaxies on
average. Observations show that brighter galaxies are more strongly clustered
than their fainter counterparts
\citep[e.g.][]{norberg01,zehavi05,pollo06,coil06}. We explicitly show in the
right panels of Fig. \ref{bimod} and in Tab. \ref{table} and Tab. \ref{table2}
that we indeed select intrinsically brighter galaxies with increasing
redshift, both in the VVDS-Deep and the SAM. The relatively constant amplitude
of the predicted correlation function suggests the absence (or a very weak)
luminosity dependence of galaxy clustering in the model \citep[see
  also][]{li07,kim09}. We study this aspect in more detail in paper II, where
we measure the clustering of galaxies with different luminosities in the model
and compare them with VVDS-Deep observations.

At $z>1.1$, Fig. \ref{wp_comp} shows that, although the overall amplitude of
the predicted galaxy correlation function is similar to that measured, the
model predicts a shallower correlation function than observed. The difference
in slope and in shape of the correlation function can be interpreted within
the framework of Halo Occupation Distribution models
\citep[e.g.][]{cooray02}. In this framework, the galaxy correlation function
is the sum of two contributions, one dominating the smaller scales that
characterises the clustering of galaxies residing in the same halo (the 1-halo
term), and a large-scale contribution, which characterises the clustering of
galaxies belonging to different haloes (the 2-halo term). At $z>1.1$, the
``bump'' of the correlation function observed on scales smaller than or of the
order of the typical halo radius ($1-2$\hmpc) in the VVDS-Deep, suggests that
these galaxies are on average hosted by relatively massive haloes, with
relatively large virial radii \citep[e.g.][]{abbas10}. The model predicts
instead a rather weak 1-halo term, which implies the presence of relatively
few satellite galaxies. Satellite galaxies are defined as the galaxies
residing within the virial radii of haloes that are not associated with their
centres. The amplitude of the 1-halo term is directly linked to the amount of
central--satellite and satellite--satellite pairs, and in turn to the
abundance of satellite galaxies in their host haloes
\citep[e.g.][]{benson00,berlind03,kravtsov04}. We have previously shown that
at $z>1.1$ the model galaxy population is dominated by blue
galaxies. Therefore, the small-scale shape of the correlation function
predicted by the SAM at these redshift, suggests that model galaxies are more
likely to be blue central galaxies in low-mass haloes rather than blue or red
satellites in more massive haloes, in contrast with what appears to be the
case in the VVDS-Deep data.

It is worth noting that the Millennium Run simulation was carried out adopting
a WMAP 1 cosmology, with a normalisation of the power spectrum of
$\sigma_8=0.9$. The most recent and accurate measurements favour instead a
lower value of $\sigma_8\simeq0.81$ \citep{komatsu09,komatsu10}. The value of
$\sigma_8$ has a non negligible influence on the amplitude of the dark matter
correlation function, and in turn to that of galaxies. To quantify the effect
on SAM clustering predictions, we convert the correlation functions in the SAM
to those expected assuming the more recent value of $\sigma_8$. To do so, we
multiply the SAM \wprp by the ratio of the non-linear projected correlation
function of mass for $\sigma_8=0.81$ to that for $\sigma_8=0.9$. We keep all
the other cosmological parameters identical. We use the \citet{smith03}
analytical prescription for the non-linear mass power spectrum, which we
Fourier transform to obtain the correlation function. With this procedure, we
assume a purely gravitational clustering evolution and do not account for a
possible dependence of galaxy bias on $\sigma_8$. This is, however, found to
be weak in simulations \citep{wang08}. By rescaling the SAM projected
correlation functions to $\sigma_8=0.81$, we lower the amplitude of the
predicted \wprp on all scales and obtain the dotted curves in
Fig. \ref{wp_comp} and Fig. \ref{color_comp}. While this tends to improve the
agreement between model predictions and observations, there are still some
differences, in particular regarding the shape of the correlation functions on
small scales.

\subsection{Colour-dependent galaxy clustering}

When selecting galaxies at $17.5<I<24$ in the redshift interval $0.2<z<2$, we
probe an evolving mix of galaxy luminosities and colours
\citep[e.g.][]{lefevre05b,ilbert05,zucca06,franzetti07}. As a consequence, the
clustering of these galaxies should reflect the average clustering of the
different galaxy sub-populations at the different redshifts. In order to
better understand the discrepancies found between the global clustering
measured from VVDS-Deep data and that predicted by the model, we compare the
clustering of two sub-sets of galaxies, red and blue ones, classified
according to their bimodal colour distribution.

\begin{figure}
  \resizebox{\hsize}{!}{\includegraphics{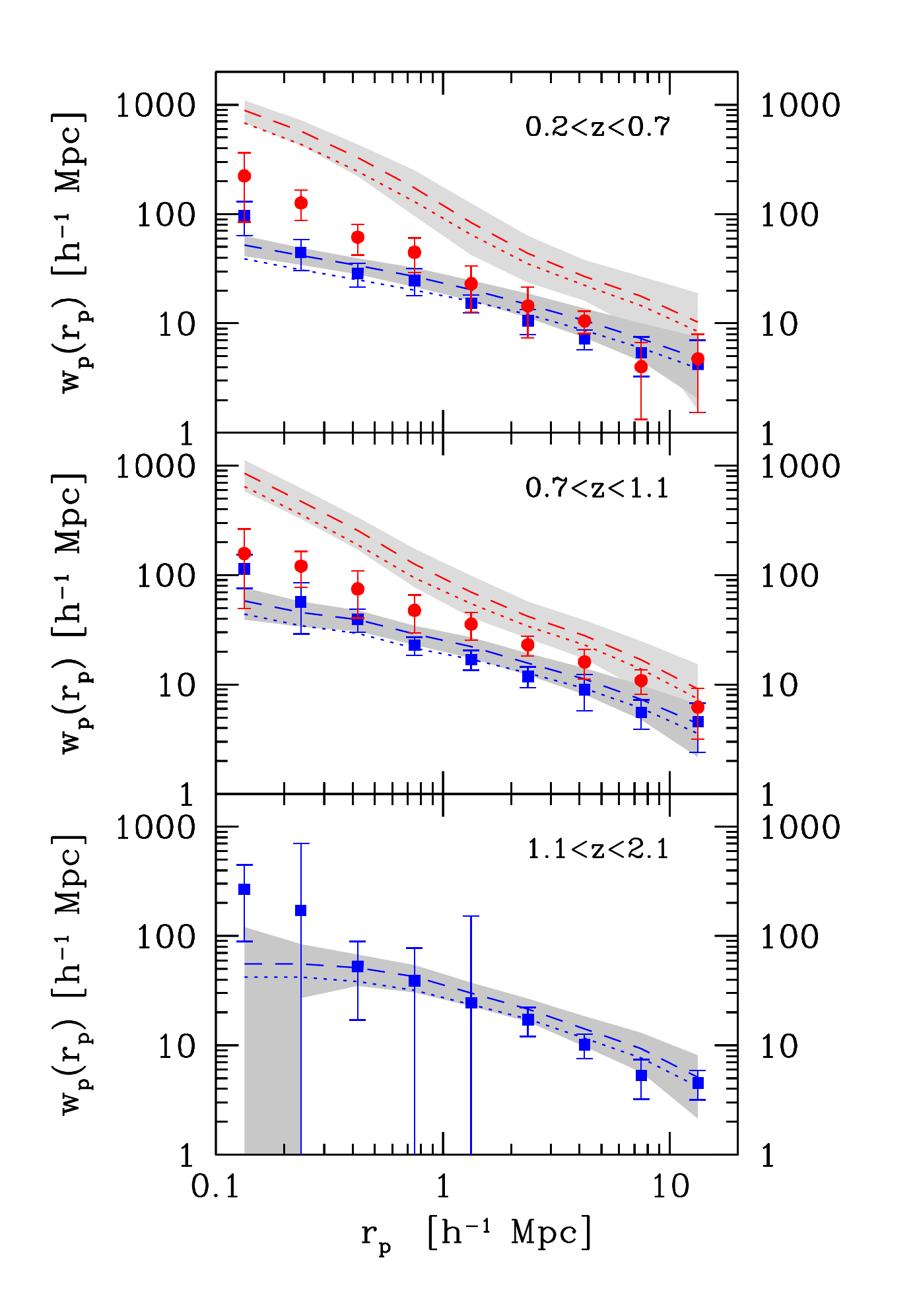}}
  \caption{Colour-dependent projected correlation function in three redshift
    intervals from $z=0.2$ to $z=2.1$, both observed in the VVDS-Deep and
    predicted by the SAM. In each panel, the filled circles (red galaxies) and
    filled squares (blue galaxies) correspond to VVDS-Deep measurements, while
    the dashed curves and associated shaded areas correspond to the mean and
    the $1\sigma$ dispersion among the \omocks. The dotted curves are the mean
    \wprp obtained by rescaling the predicted correlations function to
    $\sigma_8=0.81$ as described in Sec. \ref{sec:wpall}. In all panels, both
    for VVDS-Deep and SAM galaxies, the \wprp with higher global amplitude
    corresponds to that of red galaxies.}
  \label{color_comp}
\end{figure}

\begin{figure}
  \resizebox{\hsize}{!}{\includegraphics{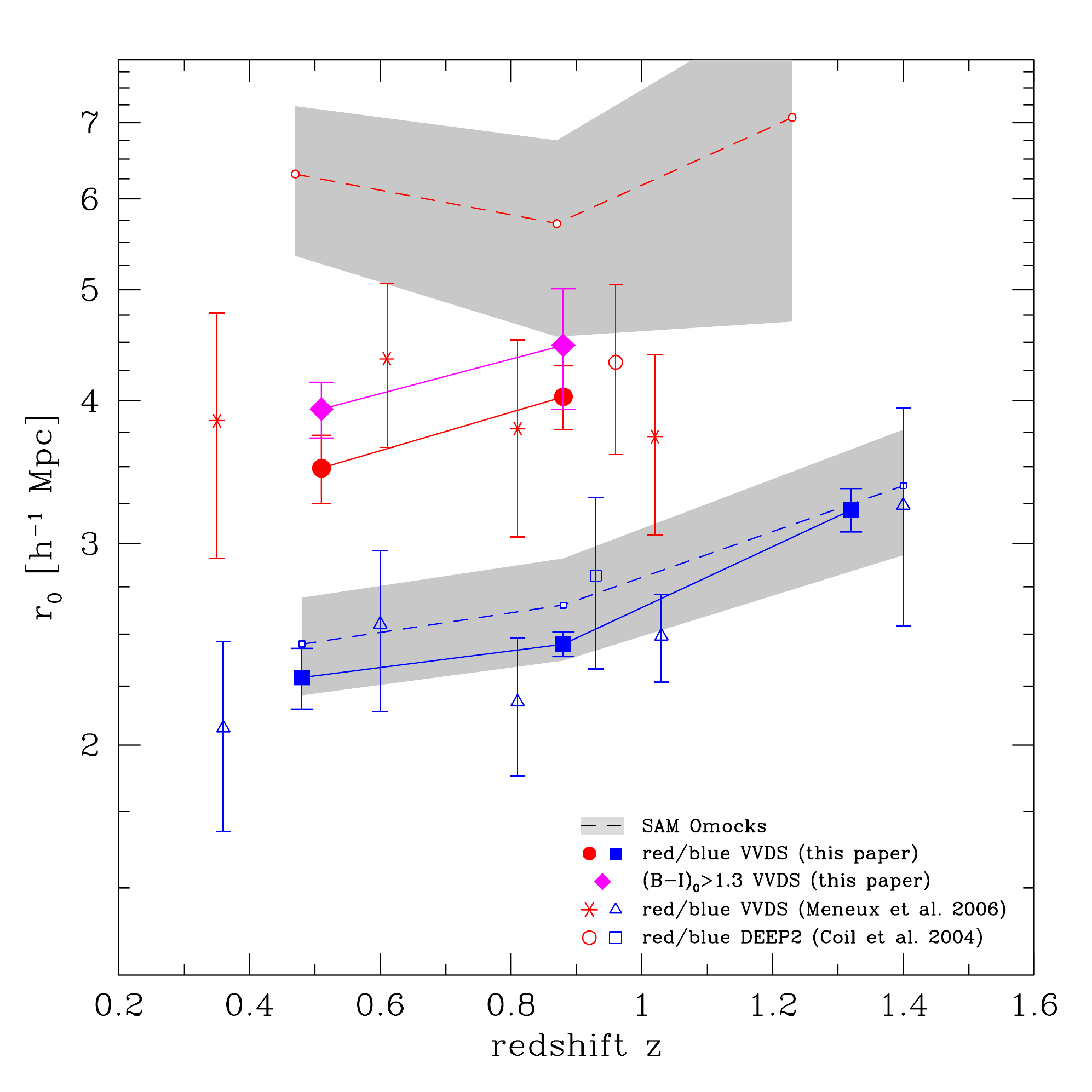}}
  \caption{Comparison of the correlation lengths of red and blue galaxies as a
    function of redshift. The filled symbols with error bars are VVDS-Deep
    measurements while the dashed curves and associated shaded areas
    correspond to the correlation lengths and errors in the \omocks. We also
    report previous VVDS-Deep measurements by \citet{meneux06} as well as
    those of \citet{coil04} from DEEP2 survey, both using different colour
    criteria. We refer the reader to the inset for the detail of the plotted
    symbols.}
  \label{rovsz_col}
\end{figure}

To keep a significant number of galaxies in each redshift interval, we used
only 3 intervals between $z=0.2$ and $z=2.1$ in this part of the analysis. We
do not consider any sample of red galaxies above $z=1.1$, as the small number
of objects prevents us from obtaining a robust \wprp measurement at these
redshifts. Fig. \ref{color_comp} and Fig. \ref{rovsz_col} show the projected
correlation functions and the correlation lengths measured in the SAM and in
the VVDS-Deep data for red and blue galaxies. The measurements from VVDS-Deep
are consistent with those obtained from the same sample by \citet{meneux06},
as well as with those obtained from the DEEP2 sample by \citet{coil04}. We
note that these two studies adopt different colour criteria to define blue and
red galaxies and that the two surveys have different observational
strategies. In particular, the DEEP2 survey selects galaxies brighter than
those in the VVDS-Deep, which explains the slightly larger observed
correlation lengths in the DEEP2 survey \citep[see][for a detailed
  discussion]{lefevre05b}.

We find that the correlation functions of blue galaxies in the model are in
quite good agreement with VVDS-Deep measurements. In contrast, red model
galaxies show a much stronger clustering on all scales.  This suggests that
the stronger clustering predicted by the SAM for the entire sample is due to
the very strong clustering of red galaxies. As for the redshift trend, both
VVDS-Deep observations and SAM predictions show a rather similar evolution
over the redshift interval probed. The correlation length of red SAM galaxies
is, however, much larger than observed and in fact similar to that observed
for extremely red objects in the real Universe \citep[e.g.,][]{daddi03}.

As explained earlier, we have used different rest-frame $B-I$ colour cuts to
define red and blue galaxies in the model and in the observations.  One may
argue that the higher clustering amplitude observed for SAM red galaxies could
be due to the redder colour cut applied to select the two populations, redder
galaxies being expected to be more strongly clustered. To test this
possibility we measure the clustering strength of red galaxies in the
VVDS-Deep using the same colour cut used in the SAM,
i.e. $(B-I)^{cut}=1.3$. In this way, we isolate in the VVDS-Deep sample red
galaxies which have the same rest-frame $B-I$ colour distribution than red
model galaxies. As shown in Fig. \ref{rovsz_col}, we find that these galaxies
in the VVDS-Deep show a higher clustering amplitude than those selected with
$B-I>0.95$. However, the SAM clustering remains significantly stronger,
demonstrating that red model galaxies are intrinsically more clustered than
observed in the VVDS-Deep.

When studying the shape of the projected correlation functions in more detail,
one finds that blue SAM galaxies are characterised by a shallower correlation
function than VVDS-Deep galaxies, in particular on small scales. Within the
HOD framework, this implies a weaker 1-halo term that can be interpreted as a
lack of blue satellite galaxies in the SAM. In contrast, red model galaxies
exhibit a correlation function which is significantly steeper and higher than
observed. Here, the very prominent 1-halo term may be due to an overabundance
of red satellite galaxies. Similarly, \citet{coil08} find an absence of
``Finger of God'' \citep[FoG,][]{jackson72} in the correlation function of
blue model galaxies at $z\simeq1$ at variance with red model galaxies, which
have a very strong FoG. The FoG effect is associated with the infall of
satellite galaxies inside haloes and its strength is related to the abundance
of satellite galaxies \citep[e.g.][]{slosar06}. These results suggest that in
the real Universe, (at least part of) the red satellites likely evolve less
rapidly than in the model, and remain in the blue tail of the colour
distribution for a longer time scale. This could adjust the different
small-scale clustering behaviours of blue and red SAM and VVDS-Deep galaxies,
but would not affect significantly the amplitude of the correlation functions.

\begin{figure}
  \resizebox{\hsize}{!}{\includegraphics{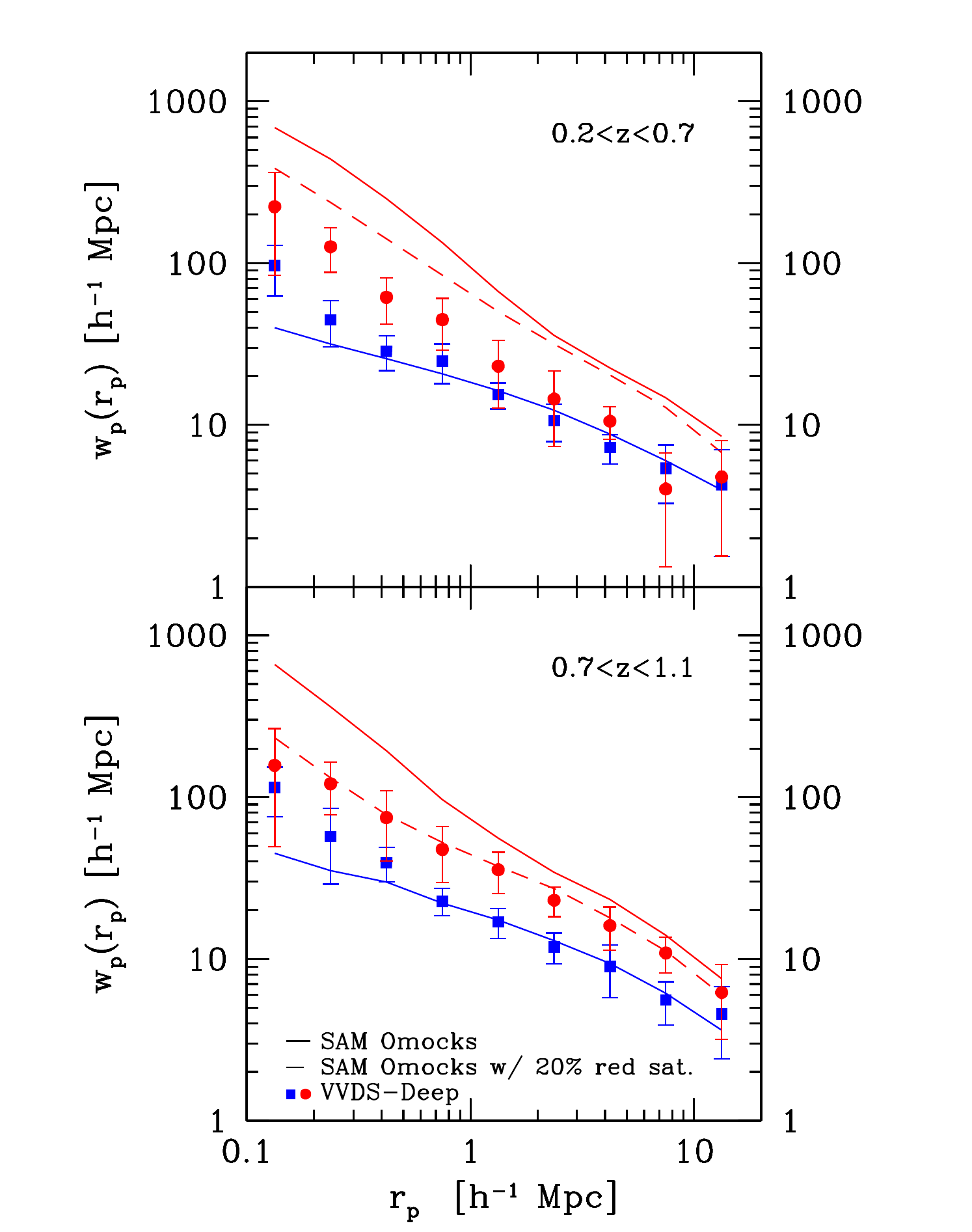}}
  \caption{Red and blue galaxy projected correlation functions in two redshift
    intervals from $z=0.2$ to $z=1.1$, both observed in the VVDS-Deep and
    predicted by the SAM (mean over \omocks samples). In each panel, the solid
    curves correspond to SAM mean predictions while the dashed ones to the
    resulting mean predictions while keeping only $20\%$ of red satellite
    galaxies. The filled circles (red galaxies) and filled squares (blue
    galaxies) correspond to VVDS-Deep measurements. Both for the VVDS-Deep and
    the SAM, the \wprp with higher amplitude corresponds to that of red
    galaxies. The SAM correlation functions are rescaled to $\sigma_8=0.81$ as
    described in Sec. \ref{sec:wpall}.}
  \label{rem_sat}
\end{figure}

To better see the impact of an overabundance of satellites on the clustering
of galaxies, we randomly remove from the SAM mock samples $80\%$ of the red
satellites. The resulting galaxies correlation functions are shown in
Fig. \ref{rem_sat}. This figure shows that, by excluding most of SAM red
satellites, the amplitude of the correlation function of red galaxies is
dramatically reduced, particularly on small scales. Model predictions obtained
excluding 80 per cent of the red satellites are in quite good agreement with
observational measurements but at $0.2<z<0.7$, where there is still a
significant difference between the amplitudes of the predicted and measured
correlation functions. Similar conclusions have been reached while comparing
the model to local measurements, e.g. \citet{li07} found that the match of the
observed clustering in the local Universe to the previous version of the
Munich model \citep{croton06} can be improved by removing $30\%$ of satellite
galaxies in the model.

\section{Summary and discussion}

We have compared some of the basic high-redshift galaxy properties as measured
in the VIMOS-VLT Deep Survey, to predictions from the Munich semi-analytical
model. For this purpose, we have constructed 100 mock samples that accurately
mimic the VVDS-Deep observational strategy. We have compared the magnitude
counts, redshift distribution, colour bimodality, and galaxy clustering for
galaxies with $17.5<I<24$, probing a broad range of cosmic epochs from z=$2$
to $z=0.2$. We have demonstrated that, in order to carry out a fair comparison
between model predictions and data, it is important to build ``observed'' mock
samples that accurately reproduce the detailed selection function and biases
of the observations.

We find that the Munich semi-analytical model reproduces reasonably well:
\begin{itemize}
\item the magnitude counts in the $u^*,~g',~r',~i',~I$ and rest-frame
  $B$ bands,
\item the shape of the redshift distribution at $z<1.8$ for $I_{AB}<24$
  galaxies, given the relatively large sample variance predicted by the model
  in the VVDS-Deep volume,
\item the global galaxy clustering at $z>0.8$,
\end{itemize}
but fails to reproduce:
\begin{itemize}
\item the magnitude counts in the $z'$ and rest-frame $I$ bands,
\item the shape of the redshift distribution at $z>2$ for $I_{AB}<24$
  galaxies,
\item the rest-frame $B-I$ colour distribution and its evolution with cosmic
  time since $z\simeq2$,
\item the clustering strength of red galaxies,
\item the detailed small-scale clustering of both red and blue galaxies.
\end{itemize}

It is important to notice that for some of the predicted galaxy properties,
there is a significant variance among different mock samples. For most of the
observational measurements discussed in this study, we find that there are a
few mock samples that are in good agreement with the data. On average,
however, models deviate from observational measurements. In particular, for
the colour distribution and the clustering of red galaxies, all mock samples
differ from the VVDS-Deep measurements, and differences are larger than
3$\sigma$. None of the mock samples is able to reproduce all the VVDS-Deep
measurements presented in this analysis, suggesting that the model failures
highlighted above are not simply due to sample variance.

The discrepancies found between model predictions and VVDS-Deep observations
extend to higher redshifts some of the model problems that have been
previously emphasised from data-model comparisons in the local
Universe. Although the blue population dominates in number density at all
redshifts, the SAM tends to produce too many relatively bright red
galaxies. As a consequence, the rest-frame $I$-band distribution is skewed
towards bright magnitudes and the rest-frame $B-I$ colour distribution towards
the red. This excess of red galaxies is dominated by satellites, giving rise
to a prominent 1-halo term in the correlation function of red model
galaxies. In addition, the SAM underpredicts the fraction of blue satellite
with respect to blue central galaxies as seen in the small-scale clustering of
blue galaxies. It is important to mention that the excess of red satellite
galaxies is not specific to the Munich semi-analytical model but is present in
most of published semi-analytical models \citep{liu10}.

The excess of red and deficit of blue satellite galaxies in semi-analytical
models are likely due to an over-efficient quenching of satellites, that
transforms too many blue galaxies to red ones over a short time scale. In the
models, star-forming blue galaxies become passive as a consequence of the
infall of a galaxy onto a larger halo, or because of AGN feedback that
suppresses star formation in massive central galaxies. An over-quenching of
satellite galaxies can be produced by a too efficient strangulation, that
instantaneously shuts off the star formation when a galaxy enters in a halo
\citep{weinmann06,font08,kang08,kimm09,fontanot09b}. It is interesting to
  note that the truncation of gas accretion in satellite galaxies, and to a
  large extend of star formation, is also found to be less abrupt in smoothed
  particle hydrodynamics simulations \citep{cattaneo07,saro10}. This would
help to explain the difference in the rest-frame colour distribution between
models and data, but cannot explain the very strong intrinsic clustering of
red galaxies. As recently pointed out by \citet{kim09}, \citet{wetzel10}, and
\citet{liu10}, the problem might lie in a poor treatment of satellite mergers
and disruption. In fact, most of current semi-analytical models, including
that used in this study, do not account for tidal stripping of satellite
galaxies \citep[but see][]{benson02,monaco07}. This can influence
significantly the predicted clustering signal, and has been shown to affect
also the predicted galaxy intrinsic colour distribution and stellar mass
function \citep{yang09}. We study these aspects in more details in paper II
where we specifically relate the predicted galaxy clustering as a function of
luminosity and colour to the halo occupation predicted by the model.

\begin{acknowledgements}
SDLT acknowledges financial support from ASI under contract ASI/COFIS/WP3110
I/026/07/0. GDL acknowledges financial support from the European Research
Council under the European Community's Seventh Framework Programme
(FP7/2007-2013)/ERC grant agreement n. 202781. AP was financed by the research
grants of the Polish Ministry of Science PBZ/MNiSW/07/2006/34A and N N203
512938. \\
This research has been developed within the framework of the VVDS
consortium.\\ 
This work has been partially supported by the CNRS-INSU and its
Programme National de Cosmologie (France), and by INAF grant PRIN-INAF
2007.\\ 
The VLT-VIMOS observations have been carried out on guaranteed time
(GTO) allocated by the European Southern Observatory (ESO) to the VIRMOS
consortium, under a contractual agreement between the Centre National de la
Recherche Scientifique of France, heading a consortium of French and Italian
institutes, and ESO, to design, manufacture and test the VIMOS instrument.
\end{acknowledgements}

\bibliographystyle{aa}
\bibliography{biblio}

\end{document}